\documentclass[11pt,a4paper]{article}

\usepackage{a4wide}
\usepackage{graphicx}
\usepackage{latexsym}
\usepackage{amssymb}
\usepackage{amstext}
\usepackage{amsgen}
\usepackage{amsxtra}
\usepackage{amsgen}
\usepackage{amsthm}

\theoremstyle{definition}

\numberwithin{equation}{section}

\begin{document}

\title{Mathematical Modelling of Polarizing GTPases \\ in Developing Axons}
\author{Natalie Emken$^*$ \and Andreas W. P\"uschel$^{\P \S}$ \and Martin Burger$^{* \S}$ }
\maketitle

\renewcommand{\thefootnote}{\fnsymbol{footnote}}
\footnotetext[1]{Institut f\"ur
Numerische und Angewandte Mathematik, Westf\"alische Wilhelms-Universit\"at (WWU) M\"unster Einsteinstr. 62, D 48149 M\"unster, Germany (\{natalie.emken,martin.burger\}@wwu.de)}
\footnotetext[5]{Institut f\"ur Molekulare Zellbiologie, WWU M\"unster, Schlossplatz 5, D 48149 M\"unster, Germany (apuschel@wwu.de)}
\footnotetext[4]{Cells in Motion Interfaculty Center (CIMIC), WWU M\"unster}
\begin{abstract}
The aim of this paper is to contribute to the basic understanding of neuronal polarization mechanisms by developing and studying a reaction-diffusion model for protein activation and inactivation. In particular we focus on a feedback loop between PI3 kinase and specific GTPases, and study its behaviour in dependence of neurite lengths. 

We find that if an ultrasensitive activation is included, the model can produce polarization at a critical length as observed in experiments. Symmetry breaking to polarization in the longer neurite is found only if active transport of a factor, in our case active PI3 kinase, is included into the model.
%{\bf Keywords: } 
%{\bf AMS Subject Classification: }  74N20, 74N25, 35K50, 35K55, 65M60.
%{\bf PACS: } 02.30.Jr, 02.30.Xx, 02.60.Lj, 81.10.Aj, 64.70.Nd
\end{abstract}

\section{Introduction}

Neuronal polarization is a key process in brain development, whose molecular origin is still far from being completely understood. Clearly, the differentiation of neurites into axons (in general only one) and several dendrites is a key mechanism for the structure and function of the brain, and thus the process is likely steered by external cues. On the other hand various lab experiments with cultured hippocampal neurons (cf. \cite{Do88,GB89}) show that a robust polarization and differentiation also occurs without external cues, driven by the activity of GTPases and other proteins. 
We refer to \cite{AB00,Ar07,Ba09,Sa96,Ta09} for detailed discussions of neuronal polarity.

During early development a neuron first extends several unspecified neurites that show random episodes of growth and retraction resulting in a constant average length. At a later stage a polarization in the distribution of several proteins occurs and usually one of the neurites is specified as the axon, which results in the stabilization of microtubuli (cf. \cite{Wi08}) and a phase of fast growth, while the other neurites grow slower and later become dendrites. The molecular origin of neuronal polarity is a spatial reorganization of proteins towards the tip of the longest neurite, which seems to appear when one of the neurites has reached a critical length (depending however on the overall configuration). In particular the sequential activity of the GTPases Rap1B, Cdc42, and Rho/Rac GTPases has been found to be of fundamental importance for the differentiation of an axon. 

The aim of this paper is to gain further understanding of neuronal polarization and symmetry-breaking as observed in cultured neurons by mathematical modelling of the GTPase signalling cascades, diffusion, and possibly transport. In particular we want to tackle the following questions:
\begin{itemize}

\item Can a feedback loop between GTPases and PI3 kinase that has been found in experiments lead to an inherent length-dependent polarization mechanism ? 

\item Is a suggested ultrasensive reaction essential for the polarization ?

\item What can cause symmetry breaking and determine the strongly preferred polarization of signaling molecules in the longest neurite ?

\end{itemize}

We want to understand these issues by modelling activation of PI3 kinase and GTPases such as Rho, Rac, Ras, Cdc42, Rap1B, which are key players in the establishment of polarity and the growth of axons (cf. \cite{Go05,Fi08,SP04,Wa06,Yo06}). Similar pathways of GTPases are found also in the polarization of other cell types (cf. \cite{Ri01,EM02}). In particular the establishment of polarity in
eukaryotic cells such as yeast, which appears to be a reasonably simple model system, has recently attracted much attention in the applied mathematics community (cf. \cite{Po01,SuNa04,Ji07,Go08,Jilkine,JiEK11}), and significant progress has been made very recently in understanding basic mechanisms as well as in the mathematical analysis of such models (cf. \cite{IS07,Mo08,MO10,Mo11}). Modelling the polarization of axons is a topic less studied (cf. \cite{Sa05}), probably also due to the higher  complexity in neurons. On the other hand the length-dependent mechanism and the preferential polarization of signaling cascades to the longer neurites is a robust process that a model should be able to explain at least qualitatively. Moreover, the changing length provides additional variability that allows to rule out certain models.

The key assumption of our modeling approach, which seems to be in good agreement with experiments (cf. e.g. \cite{GB89}) is that none of the neurites is predetermined to become the axon, and there is no polarization in neurite lengths without polarization in signaling proteins, more precisely GTPases. As described above, we consider the growth (and retraction) in the initial stage as random, and thus there can be some polarization in neurite length. However, without polarization in molecular distributions it seems quite unreasonable that a neurite at a certain length can become stable and subsequently differentiates to the axon. Thus, we ask whether mathematical modelling of GTPase activation can explain the stabilization of the longest neurite when it has reached a critical length by stochastic growth. 

As we want to focus on the polarization of GTPase distributions it is natural to set up a reaction-diffusion model to study spatial variation. Noticing that overall growth is slow, it is natural to assume that the reaction-diffusion process reaches equilibrium at fixed length, in particular if the equilibria are close also after an incremental growth step. The latter applies in particular for stationary concentrations being spatially homogeneous and thus we can argue that polarization will only appear if at a certain critical length the homogeneous stationary solution becomes unstable and a polarized solution showing a peak on one end appears. There are several suggestions of extracellular and stochastic processes that influence polarity (cf. \cite{Ar07,Ba09,Ta09}), which can be understood as small perturbations of the homogeneous state. Thus, we consider the modelling of GTPase activation on fixed domains, but investigate the influence of different domain sizes. In particular we expect the homogeneous stationary state to become unstable at a critical length, which we interpret as length-dependent polarization.

The remainder of the paper is organized as follows: In Section 2 we will develop a basic mathematical model of activation mechanisms in developing neurites, which is further simplified in Section 3 and investigated numerically. In Section 4 we discuss the extension to an appropriate setup for a neuron with two neurites and investigate the possibility of symmetry-breaking. Finally we provide further analytical insight via a minimal model in Section 5 and conclude in Section 6.

\section{A Model for Activation of GTPases in Neurites}

In the following we derive a reaction-diffusion model for the polarization in neurites starting with a general approach to the modelling of GTPase activation and then proceeding to a specific signal pathway, which results in a system of reaction-diffusion equations.

\subsection{General Aspects in Modelling GTPases}  \label{GTPasenModell}

As mentioned above GTPases, appearing in cytosolic and membrane-bound forms, are of central importance for regulating neuronal polarization via feedback-loops and signaling growth. We therefore start by discussing the general modelling of GTPases first before turning to the specific signalling cascade. We follow the basic approach of \cite{Ji07,Jilkine} and first detail a submodel for a single GTPase (ignoring for the moment the signal pathways between different molecules), composed of three states the GTPase can switch to.  There is an inactive cytosolic form (bound to GDI), an inactive membrane-bound form (bound to GDP), and an active membrane-bound form (bound to GTP). 
For the concentrations we use the following notations:
\begin{itemize}
\item \begin{math} G_a \end{math} denotes the membrane-bound active form
\item \begin{math} G_m \end{math} denotes the membrane-bound inactive Form
\item \begin{math} G_c \end{math} denotes the cytosolic inactive form
\end{itemize}

\begin{figure}[h]
\begin{center}
\includegraphics[height=5cm]{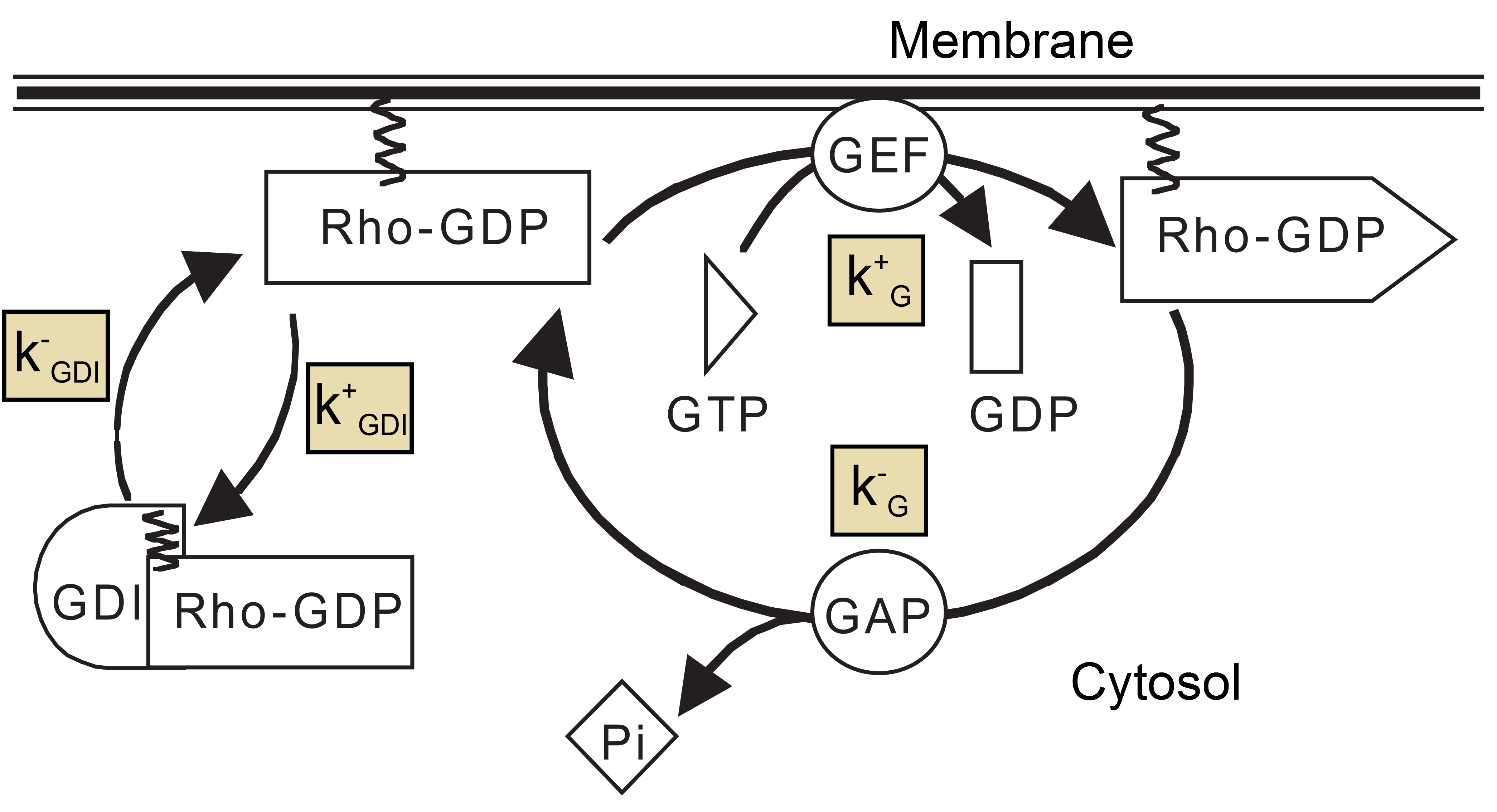}
\caption{\label{ReakGTPasen}GTPase reaction scheme (following \cite{EM02}). The GDI-bound form diffuses in the cytosol and associates to the membrane with rate
 $k^-_{GDI}$. On the other hand the membrane-bound form is released and bound to GDI with rate $k^+_{GDI}$. }
\end{center}
\end{figure}

We base the modelling on similar assumptions as \cite{Ji07}, motivated by experimental observations:
\begin{enumerate}
\item  Neuronal polarization is sufficiently fast, also compared to growth. Thus we consider models on fixed geometries (however focusing on the impact on different neurite lengths) and assume that Rho-proteins are neither synthesized nor degradated at the relevant time scale. We only model the change between different forms.

\item Each Rho-protein has spectific rates of activation and inactivation. These processes are indirect and controlled by GEFs and GAPs, respectively (see Figure  \ref{ReakGTPasen}).

\item The amount of GDI in the developing neuron is sufficiently high such that it does not become a limiting factor, and the change between the cytosolic and membrane-bound form is sufficiently fast. The latter is not crucial, but will allow to reduce the model to two forms, making the model easier to treat. 

\item Cytosolic forms diffuse much faster than membrane-bound forms (by a factor of around 100, cf. \cite{Po01}). Again this is not crucial for the basic modelling, but will become crucial for forming a Turing-type instability: the concentration of cytosolic forms will be rather homogeneous, while active forms will polarize at the membrane. 

\item Since Rho-Proteins have almost equal sizes, we will assume that the diffusion coefficients of different proteins in the active respectively inactive form are the same. This assumption is not crucial for the modelling and easily changed in the equations, but it appears convenient for numerical simulations. 
\end{enumerate}

Let us denote the interior of the neurite as $\Omega$ and the membrane as $\Gamma \subset \partial \Omega$. The cytosolic concentration is modeled by a bulk diffusion equation
\begin{equation}
	\partial_t G_c = D_c \Delta G_c \qquad \text{in } \Omega,
\end{equation}
with $\Delta$ denoting the Laplace operator. The membrane-bound concentrations are modeled by reaction-diffusion equations on the surface
\begin{align}
\partial_t G_a &= D_m \Delta_\Gamma G_a + R_{am}(G_a,G_m) &\text{on } \Gamma \\
\partial_t G_m &= D_m \Delta_\Gamma G_m - R_{am}(G_a,G_m) + R_{mc}(G_m,G_c) &\text{on } \Gamma,
\end{align} 
where $\Delta_\Gamma$ denotes the Laplace-Beltrami operator on $\Gamma$. The reactions (activation and inactivation)  between the two membrane-bound forms are given by
\begin{equation}
	R_{am}(G_a,G_m) = -k^-_{G} \text{GAP}_G G_a + k^+_G \text{GEF}_G G_m,
\end{equation}
with the activation rate $k^+_G$ and inactivation rate $k^-_{G}$, 
and the reactions (exchange) between the inactive forms are given by 
\begin{equation}
	R_{mc}(G_m,G_c) = -k^-_{GDI} G_m+ k^+_{GDI} G_c,
\end{equation}
with the GDI-controlled rates $k^+_{GDI}$ und $k^-_{GDI}$  (cf. Figure \ref{ReakGTPasen}). Finally, the boundary condition for $G_c$ is given by
\begin{equation}
	D_c \nabla G_c \cdot n = - R_{mc}(G_m,G_c) \qquad \text{on } \Gamma.
\end{equation}

We can further simplify the system based on the assumption of a fast exchange between the two inactive form, such that $R_{mc} = 0$ on $\Gamma$, i.e. 
\begin{equation}
0=k^-_{GDI}G_m-k^+_{GDI}G_c\qquad \Leftrightarrow \qquad %G_m=\frac{k^+_{GDI}}{k^-_{GDI}}G_c, \qquad
G_c=\frac{k^-_{GDI}}{k^+_{GDI}}G_m.  \label{eq:GmGcrelation}
\end{equation}
Finally, taking into account the elongated form of neurites, it is natural to reduce to a one-dimensional system. For this sake assume that $x$ is the longitudinal axis of a neurite and let $P_\xi$ denote the two-dimensional plane where $x=\xi$. Now we define effective concentrations of active and inactive forms
\begin{eqnarray}
u_a(x,t) &=& \int_{P_x \cap \Gamma} G_a(x,y,z,t)~ds(y,z),  \\
u_i(x,t) &=& \int_{P_x \cap \Gamma} G_m(x,y,z,t)~ds(y,z) + \int \int_{P_x \cap \Omega} G_i(x,y,z,t)~dy~dz .
\end{eqnarray}
Since we assume cross-sections to be small relative to the length, we can assume equilibration by diffusion in the $y$ and $z$ directions at a time scale relevant for longitudinal diffusion and reaction, such that we can well approximate all concentrations as independent of $y$ and $z$. 
In particular we have with \eqref{eq:GmGcrelation}
\begin{equation}
	u_a \approx \vert P_x \cap \Gamma \vert  G_a, \qquad 
	u_i \approx (	 \vert P_x \cap \Omega \vert ~\frac{k^-_{GDI}}{k^+_{GDI}} + \vert P_x \cap \Gamma \vert)  G_m := \lambda G_m.
\end{equation}

Thus, we find
$$
	\partial_t u_a = \int_{P_x \cap \Gamma} \partial_t G_a~ds(y,z) \approx \vert P_x \cap \Gamma \vert ( D_m  \partial_{xx} G_a  + R_{am}(G_a,G_m) ) \approx D_m \partial_{xx} u_a  + R_{am}(u_a,\lambda^{-1} \vert P_x \cap \Gamma \vert  u_i)
$$
and
\begin{eqnarray*}
\partial_t u_i &=& \int_{P_x \cap \Gamma} \partial_t G_m ~ds(y,z) + \int \int_{P_x \cap \Omega} \partial_t G_i~dy~dz \\ &\approx& 
 \vert P_x \cap \Gamma \vert( D_m \partial_{xx}  G_m - R_{am}(G_a,G_m))+ D_c \vert P_x \cap \Omega \vert  \partial_{xx}  G_c  \\
   &\approx& D_{mc} \partial_{xx} u_i - R_{am}(u_a,\lambda^{-1} \vert P_x \cap \Gamma \vert  u_i),
\end{eqnarray*}
with the effective diffusion coefficient 
\begin{equation}
	D_{mc} = \lambda^{-1} \vert P_x \cap \Gamma \vert D_m + \lambda^{-1} \vert P_x \cap \Omega \vert D_c \frac{k^-_{GDI}}{k^+_{GDI}}.
\end{equation}
Thus, with redefining $\tilde k^+_G = \lambda^{-1} \vert P_x \cap \Gamma \vert~ k^+_G$, we end up with the system
\begin{eqnarray}
\partial_t u_a &=& D_m \partial_{xx} u_a -k^-_{G} \text{GAP}_G u_a + \tilde k^+_G \text{GEF}_G u_i, \\
\partial_t u_i &=& D_{mc} \partial_{xx} u_i+ k^-_{G} \text{GAP}_G u_a - \tilde k^+_G \text{GEF}_G u_i 
\end{eqnarray} 
for the activated and inactivated forms of the GTPases. Note that under typical parameter values we still expect $D_{mc}$ to be much larger than $D_m$ due to its dependence on $D_c$.

\subsection{Modelling the PI3-Kinase Signal Pathway}

In the following we try to model the most important signal pathways leading to neuronal polarization. We mainly follow signal cascades described in \cite{Ar07,Fi08,Yo06,Ta09}. Motivated by polarization models for eukaryotic cells, we particularly look for a feedback mechanism, which can lead to bistability or Turing-type instability. Such a feedback loop is known for the activation of PI3 kinases, which is discussed as a possible source for neuronal polarity \cite{Ar07,Fi08}. 

\begin{figure}[h]
\begin{center}
\includegraphics[width=\textwidth]{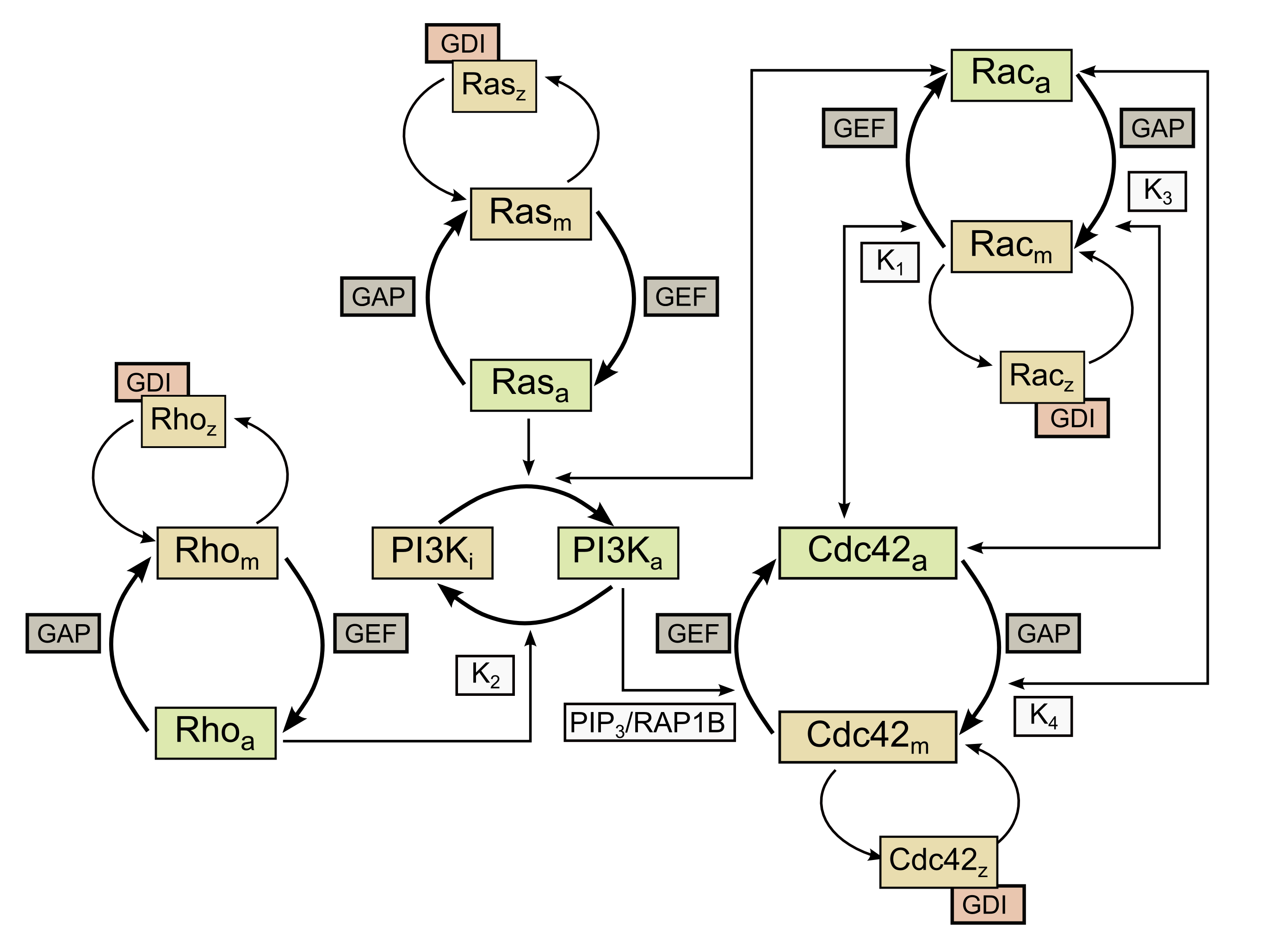}
\caption{\label{Reaktionsschema}Signalling scheme of GTPases involved in our model including the feedback loop PI3kinase - Cdc42 - Rac.}
% sowie der induzierten Rückkopplungsschleife. Während GTP-gebundenes Ras die Aktivierung der PI3 Kinase bewirkt, führt aktives Rho über die Phosphatase PTEN ($K_2$) zu ihrer Inaktivierung. Die aktive PI3 Kinase ist über ihr Produkt PIP$_3$ und über RAP1B in der Lage, den GDP/GTP-Austausch der GTPase Cdc42 zu katalysieren. Diese bewirkt schließlich den GDP/GTP-Austausch der GTPase Rac und führt wiederum zur Aktivierung der PI3 Kinase.}
\end{center}
\end{figure}

During polarization PI3kinases generate a feedback loop in a process involving their product PIP$_3$, the GTPases Ras, Rap1B, Cdc42, Rho, and Rac, as well as complexes such as PAR3/PAR6/aPKC. In order to keep the modelling at a reasonable complexity, we will not go into details of the biochemistry, but just assume some activity of canonical complexes involved in all the inhibitions and activations. The basic signalling pathway is sketched in Figure \ref{Reaktionsschema}, where the $K_i$ denote complexes involved and the indices $a$, $i$, $c$ denote active, inactive membrane-bound, and cytosolic form, respectively. As demonstrated in the previous section, we will reduce all GTPases to an active and an inactive form. 

The GDP/GTP exchange of Rac is controlled by Cdc42 via the PAR3/PAR6/aPKC complex and recruitment of specific GEFs (like TIAM and STEF).  We thus use a complex $K_1$ for the activation of Rac in our model. 
PI3 kinase is activated by the active forms of Ras and Rac, and inhibited by PTEN, which is controlled by Rho. We thus assume a complex $K_2$, which contributes to inactivation of PI3 kinase.
Moreover, PI3kinase indirectly activates Cdc42 via PIP$3$ and Rap1B (cf. \cite{SP04}).  In addition to those factors specific GAPs for Cdc42 and Rac can intercept the feedback loop. We will denote them as complexes $K_3$ und $K_4$.

We will model the above reactions simply via the law of mass action, except one, namely the activation of PI3 kinase by Rac, for which we use a square law (alternatively Hill-type activation as e.g. in \cite{Mo11} are possible). Frequently ultrasensitive reactions, i.e. significantly steeper than the shape of Michaelis-Menten kinetics, are supposed. Indeed there is an experimental observation of ultrasensitivity of Rap1B against signals of different receptors reported in \cite{Li09}.
Since the Rap1B signalling pathways is related to PI3 kinase, we assume its ultrasensitive activation by Rac in our model. However, we think that also other ultrasensitive reactions in the feedback loop PI3 kinase - Cdc42 - Rac would have a similar effect - a detailed determination of those remains a challenge for future experiments. On the other hand the existence of an ultrasensitive reaction seems crucial as we shall show in a minimal model in Section \ref{sec:minimalmodel}.

The above considerations yield the following reaction-diffusion model for the PI3 kinase concentration
\begin{align}
{\partial_t PI3K_a}=&k^+_6Ras_aPI3K_i+k^+_7Rac_a^2PI3K_i-k^+_{5}K_2PI3K_a%\\ \notag & 
+D_{PI3K_{a}} {\partial_{xx} PI3K_a}, \label{eqn:Sys1a}\\
{\partial_t PI3K_i}=&-k^+_6Ras_aPI3K_i-k^+_7Rac_a^2PI3K_i+k^+_{5}K_2PI3K_a%\\ \notag & 
+D_{PI3K_{i}}\partial_{xx} PI3K_i  .
\end{align}
Using the above reaction scheme in addition to the general submodel for GTPases from the previous section, we obtain a system of eight differential equations for the active and inactive forms of  Ras, Cdc42, Rac, and Rho: 
\begin{align}
\partial_t Ras_a =& k^+_{Ras} Ras_i-k^-_{Ras} Ras_a +D_m \partial_{xx} Ras_a ,\\
\partial_t Ras_i =& -k^+_{Ras} Ras_i+k^-_{Ras} Ras_a + D_{mc}\partial_{xx} Ras_i ,
\end{align}
\begin{align}
\partial_t Cdc_a=&k^+_{Cdc} Cdc_i+k^+_{8}  PI3K_aCdc_i-k^-_{Cdc}Cdc_a-k^-_{9}K_4Cdc_a\\ \notag &-k^+_{1} Cdc_a+ k^-_{1}K_1 -k^+_3Cdc_a+k^-_{3}K_3+D_m \partial_{xx} Cdc_a ,\\
\partial_t Cdc_i =&-k^+_{Cdc} Cdc_i-k^+_{8}  PI3K_aCdc_i+k^-_{Cdc}Cdc_a\\ \notag &+k^-_{9}K_4Cdc_a + D_{mc} \partial_{xx} Cdc_i ,
\end{align}
\begin{align}
\partial_t Rac_a = &k^+_{Rac} Rac_i +k^+_{10}  K_1Rac_i-k^-_{Rac}Rac_a-k^-_{11}K_3Rac_a\\ \notag &-k^+_4 Rac_a+ k^-_{4}K_4 +D_m \partial_{xx} Rac_a ,\\
\partial_t Rac_i = &-k^+_{Rac} Rac_i -k^+_{10}  K_1 Rac_i+k^-_{Rac}Rac_a +k^-_{11}K_3Rac_a \\ \notag & + D_{mc}\partial_{xx} Rac_i ,
\end{align}
\begin{align}
\partial_t Rho_a = &k^+_{Rho} Rho_i-k^-_{Rho} Rho_a -k^+_2 Rho_a+ k^-_{2}K_2 +D_m \partial_{xx} Rho_a ,\\
\partial_t Rho_i = &-k^+_{Rho} Rho_i+k^-_{Rho} Rho_a + D_{mc} \partial_{xx} Rho_i ,
\end{align}
Finally we obtain model equations for the different complexes, for which we ignore diffusion:
\begin{flalign*}
\partial_t K_1 =&k^+_1Cdc_a-k^-_{1}K_1,\qquad 
\partial_t K_2 =k^+_2Rho_a-k^-_{2}K_2,\\
\partial_t K_3 =&k^+_3Cdc_a-k^-_{3}K_3,\qquad
\partial_t K_4 =k^+_4Rac_a-k^-_{4}K_4.\label{eqn:Sys1b} 
\end{flalign*}

We mention that the model can be seen as a mass-conserved reaction-diffusion system similar to those proposed for eukaryotic cells. In particular the following conservation 
properties hold in the absence of diffusion:
\begin{align}
&\partial_t(Ras_a+Ras_i)=0,\\
&\partial_t(Cdc_a+Cdc_i+K_1+K_3)=0,\\
&\partial_t(Rac_a+Rac_i+K_4)=0,\\
&\partial_t(Rho_a+Rho_i+K_2)=0,\\
&\partial_t(PI3K_a+PI3K_i)=0.
\end{align}
We mention that the result of our modelling is a system of 14 equations, which is still difficult to handle. We will therefore try to further reduce it to a system of six equations for the active and inactive forms of PI3 kinase, Cdc42, and Rac, whose feedback loop shall be the key aspect for polarization.

\section{A Six-Equations Model}

In the following we shall further reduce the system. First of all we assume that the reactions via all complexes are fast at the relevant time scales, which yields 
\begin{align*}
K_1=&\frac{k^+_1}{k^-_{1}}Cdc_a:=\kappa_1 Cdc_a,\qquad
K_2=\frac{k^+_2}{k^-_{2}}Rho_a:=\kappa_2 Rho_a,\\
K_3=&\frac{k^+_3}{k^-_{3}}Cdc_a:=\kappa_3 Cdc_a,\qquad
K_4=\frac{k^+_4}{k^-_{4}}Rac_a:=\kappa_4 Rac_a.
\end{align*}
Moreover, we try to eliminate the equations for the GTPases Ras and Rho, since they act on PI3 kinase, but are not influenced by the components of the feedback loop - we therefore assume they equilibrate and obtain the relations
\begin{align}
Ras_a&=\frac{k^+_{Ras}}{k^-_{Ras}} Ras_i\\
Rho_a&=\frac{k^+_{Ras}}{k^-_{Rho}} Rho_i.
\end{align}
Further approximating the inactive concentration as homogeneous and using the conservation properties we find
\begin{align}
Ras_a&=\frac{\frac{k^+_{Ras}}{k^-_{Ras}}}{1+\frac{k^+_{Ras}}{k^-_{Ras}}} Ras_{*}:=\gamma_1 Ras_{*},\\
Rho_a&=\frac{\frac{k^+_{Ras}}{k^-_{Rho}} }{1+\frac{k^+_{Ras}}{k^-_{Rho}}+\kappa_2 \frac{k^+_{Ras} }{k^-_{Rho}}}Rho_{*}:=\gamma_2 Rho_{*},
\end{align}
where the index $*$ denotes the total concentrations in the system, which we assume fixed. 

Inserting those simplifications into (\ref{eqn:Sys1a})-(\ref{eqn:Sys1b}) we finally obtain  a system of six reaction-diffusion equations for the concentrations of PI3 kinase and the GTPases Cdc42 and Rac:
\begin{flalign}
\label{eqn:Final1}\partial_t Cdc_a =&k^+_{Cdc} Cdc_i+k^+_{8}  PI3K_aCdc_i-k^-_{Cdc}Cdc_a\\ \notag &-k^-_{9}\kappa_4 Rac_aCdc_a +D_m\partial_{xx} Cdc_a ,\\
\partial_t Cdc_i =&-k^+_{Cdc} Cdc_i -k^+_{8}  PI3K_a Cdc_i+k^-_{Cdc}Cdc_a\\ \notag &+k^-_{9}\kappa_4 Rac_aCdc_a + D_{mc}\partial_{xx} Cdc_i , \\
%\end{flalign}
%\begin{flalign}
\partial_t Rac_a = &k^+_{Rac} Rac_i +k^+_{10}  \kappa_1 Cdc_aRac_i -k^-_{Rac}Rac_a\\ \notag &-k^-_{11}\kappa_3 Cdc_aRac_a+D_m \partial_{xx} Rac_a ,\\
\partial_t Rac_i = &-k^+_{Rac} Rac_i-k^+_{10}  \kappa_1 Cdc_aRac_i+k^-_{Rac}Rac_a\\ \notag &+k^-_{11}\kappa_3 Cdc_aRac_a+ D_{mc}\partial_{xx} Rac_i ,\\
\partial_t PI3K_a =&-k_{5}\kappa_2 \gamma_1 Rho_{*} PI3K_a+k_6\gamma_2 Ras_{*} PI3K_i\\ \notag &+k_7Rac_a^2 PI3K_a+D_{PI3K_a}\partial_{xx} PI3K_a ,\\
\label{eqn:Final2}  \partial_t PI3K_i =&+k_{5}\kappa_2 \gamma_1 Rho_{*} PI3K_a-k_6\gamma_2 Ras_{*} PI3K_i\\ \notag &-k_7Rac_a^2 PI3K_i+D_{PI3K_i}\partial_{xx} PI3K_i.
\end{flalign}

\subsection{Dimensionless Variables and Scaling}

In order to eliminate the dependence on dimensions and in particular to rescale the domain to unit size, which simplifies the systematic investigation of length-dependent polarization, we perform a scaling of the variables in the system \eqref{eqn:Final1}-\eqref{eqn:Final2}. We denote the dimensional variables used in the previous section by $\tilde x$ and $\tilde t$ and introduce new space and time variables and new concentrations via
$$\begin{array}{rclrclrcl}
t&=&\frac{\tilde t}{\tau}, \qquad &x &=&\frac{\tilde x}{L}, \qquad &\epsilon &=& \frac{L_0}{L}\\
u_1 &=&Cdc_a L, \qquad &u_3 &=& Rac_a L,\qquad &u_5 &=& PI3K_a L \\
u_2 &=&Cdc_i L, \qquad &u_4 &=& Rac_i L,\qquad &u_6 &=& PI3K_i L \\
\end{array}$$
where $\tau$ is a suitable time scale and $L$ is the total length of the system (to be varied in the following). $L_0$ ia an arbitrary but fixed reference length used to introduce the dimensionless parameter $\epsilon$ and thus formulate the question of length-dependent polarization in terms of $\epsilon$. For $\epsilon$ large (small length) we do not expect polarity, thus there should be a homogeneous solution ${\bf u}=(u_1,u_2,u_3,u_4,u_5,u_6)$. At critical $\epsilon$, the homogeneous solutions should become unstable and different solutions showing maxima of the active concentrations ($u_i$ for $i$ odd) at the boundary should appear. 

The scaled version of 
(\ref{eqn:Final1})-(\ref{eqn:Final2}) is given by
\begin{align}
\partial_t u_1 =&\alpha_1 u_2+\epsilon \alpha_2 u_{5} u_2 -\alpha_3 u_1 -\epsilon \alpha_4 u_3u_1+d_1\epsilon^2\partial_{xx} u_1,\label{eqn:Sys2a}\\
\partial_t u_2 =&-\alpha_1 u_2-\epsilon \alpha_2 u_{5} u_2 +\alpha_3 u_1 +\epsilon \alpha_4 u_3 u_1 + \epsilon^2 \partial_{xx} u_2,\\
\partial_t u_3 =&\alpha_5 u_4+\epsilon \alpha_6 u_1 u_4-\alpha_{7} u_3 -\epsilon \alpha_{8} u_1 u_3+d_1\epsilon^2 \partial_{xx} u_3,\\
\partial_t u_4 =&-\alpha_5 u_4-\epsilon \alpha_6 u_1 u_4+\alpha_{7} u_3 +\epsilon \alpha_{8} u_1 u_3+\epsilon^2 \partial_{xx} u_4,\\
\partial_t u_5 =&-\alpha_{9}u_5+ \alpha_{10}u_6+\epsilon^2 \alpha_{11} u_3^2 u_{6}+d_2\epsilon^2 \partial_{xx} u_5,\\
\partial_t u_{6} =&\alpha_{9}u_5- \alpha_{10}u_6-\epsilon^2 \alpha_{11} u_3^2 u_{6}+d_3\epsilon^2 \partial_{xx} u_{6},\label{eqn:Sys2b}
\end{align} 
with the new parameters given in Table \ref{Parameter2}.

\begin{table}
\center
{%\footnotesize 
\begin{tabular*}{10cm}[]{ll}%\toprule
%\textbf{Parameter}  & \textbf{Parameter} \\% \midrule
$\alpha_1= k^+_{Cdc}L_0^2/D_{mc}$ &
$\alpha_2= k^+_8   L_0/D_{mc}$\\ $\alpha_3=k^-_{Cdc} L_0^2/D_{mc}$ & 
$\alpha_4=k^-_9\kappa_4L_0/D_{m}$\\ $\alpha_5=k^+_{Rac}L_0^2/D_{mc}$ &
$\alpha_6=k^+_{10}  \kappa_1 L_0/D_{mc}$\\
$\alpha_7=k^-_{Rac}L_0^2/D_{mc}$ &
$\alpha_8=k^-_{11} \kappa_3 L_0/D_{mc}$\\ $\alpha_{9}=k_5 \kappa_2 \gamma_1 Rho_{*} L_0^2/D_{mc}$ &
$\alpha_{10}=k_6 \gamma_2 Ras_{*} L_0^2/D_{mc}$ \\
$\alpha_{11}=k_7/D_{mc}$&
$d_1=D_m/D_{mc}$\\ $d_2=D_{PI3K_a}/D_{mc}$&
$d_3=D_{PI3K_i}/D_{mc}$\\ $\tau=L_0^2/D_{m}$&
$\epsilon=L_0/L$ \\ %\bottomrule
\end{tabular*}}
\caption{\label{Parameter2}Overview of dimensionsless parameters in (\ref{eqn:Sys2a})-(\ref{eqn:Sys2b})}
\end{table}

We use this model now on the unit interval with no-flux boundary conditions, i.e. 
\begin{equation}
\left. \partial_x u_i \right|_{x=0,1}=0.\label{eqn:RandModell}
\end{equation}
Note that the system has three volume-conserving subsystems, we have
\begin{equation}
\partial_t \int_0^1 ( u_i + u_{i+1}) dx = 0, \qquad \text{for }i=1,3,5.
\end{equation} 

Of course it would be an ideal case to obtain the parameters in the model from experimental observations, which are currently not available in most cases however. Based on data on protein diffusion used in \cite{Po01} we use a factor $100$ in the diffusion coefficients between the cytosolic and membrane-bound forms of GTPases. Their values for cytosolic diffusion coefficients are around $10-50 \mu m^2/s^{-1}$ and for membrane diffusion coefficients around $0.1 \mu m^2/s^{-1}$, which we use for our simulations.
For PI3 kinase we use the value of $0.1 \mu m^2/s^{-1}$, since it is still bound to the membrane via receptors. Moreover we assume cytosolic diffusion for the inactive form and thus a diffusion speed of $10 \mu m/s^{-1}$, motivated by its association with inactive Rap1B. Rates of activation and inactivation cannot be found in literature, so we use rough estimates leading to reasonable time scales. The used dimensionless values are given in Table \ref{Tab:homPa}.

\subsection{Numerical Results}

In the following we present results of numerical simulations of the system (\ref{eqn:Sys2a})-(\ref{eqn:Sys2b}) on the unit interval with boundary conditions (\ref{eqn:RandModell}). All results were computed in MATLAB with the \texttt{pdepe} tool. To check that there are no discretization artefacts and that computed equilibria are robust we also used an independent upwind finite difference implementation with backward Euler time discretization, which led to the same results up to numerical errors and are not displayed here. 

The initial value in all simulations is a perturbation of homogeneous states for all concentrations. Both cosine-type and random perturbations were used, which lead to the same results except that the direction of polarity is predefined by the sign of the cosine perturbation, while random perturbations lead to random choice of direction.

\begin{table}
\begin{center}
\begin{tabular}{llcll}
{\bf Parameter} &{\bf Value}& \quad &  {\bf Parameter} &{\bf Value} \\
$\alpha_1$ & $0.001$ & & $\alpha_8$ & $0.01$ \\
$\alpha_2$ & $0.5$ & & $\alpha_9$ & $0.1$ \\
$\alpha_3$ & $0.01$  & & $\alpha_{10}$ & $5$\\
$\alpha_4$ & $0.01$  & & $\alpha_{11}$ & $0.001$ \\
$\alpha_5$ & $0.001$  & & $d_1$ & $0.01$\\
$\alpha_6$ & $1$  & & $d_2$ & $0.01$ \\
$\alpha_7$ & $0.01$   & & $d_3$ & $0.01$ \\
\end{tabular}
\caption{\label{Tab:homPa}Dimensionless parameter values used for numerical simulations}
\end{center}
\end{table}

\begin{figure}[h]
\begin{center}
\includegraphics[width=0.3\textwidth]{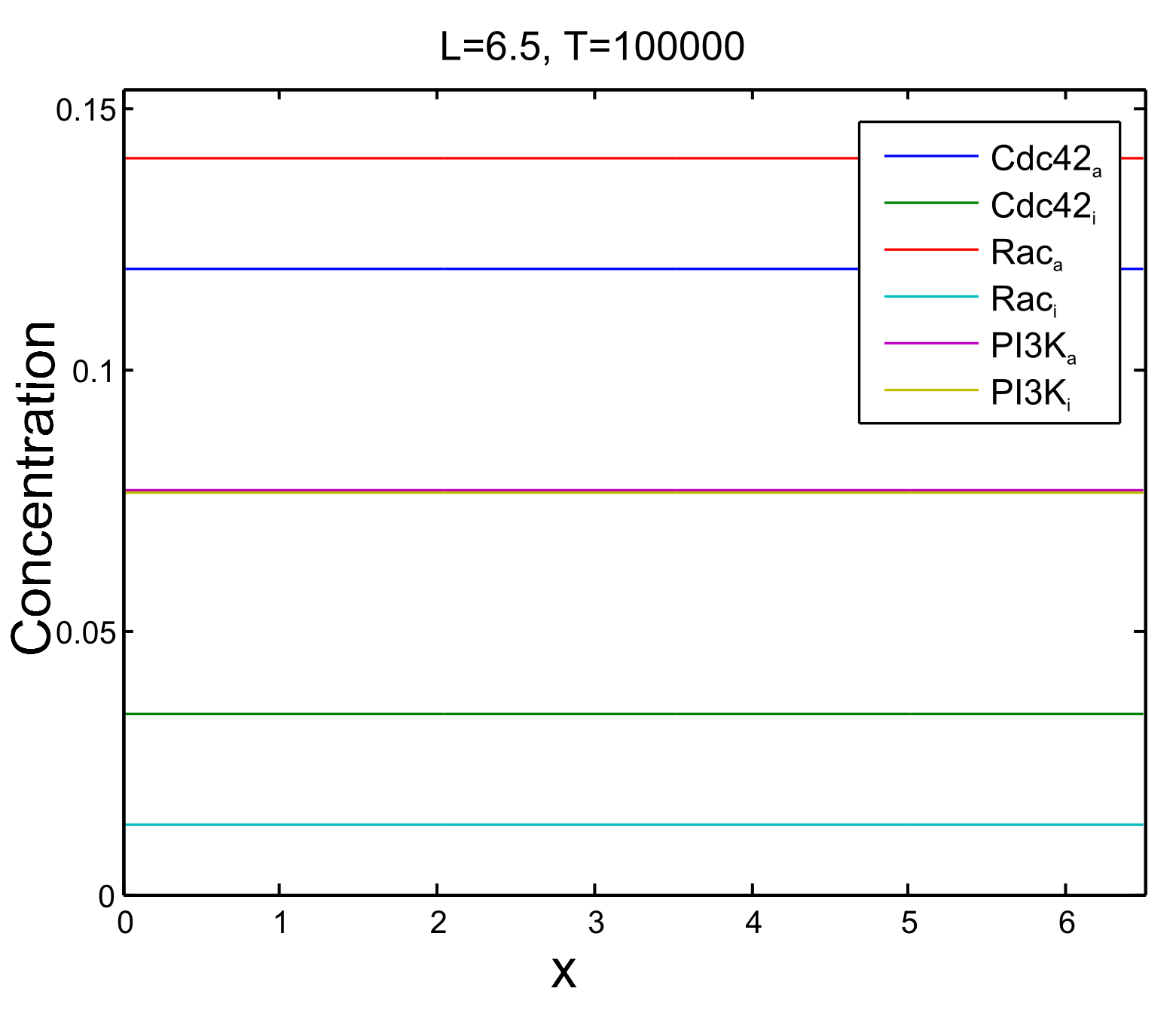}
\includegraphics[width=0.3\textwidth]{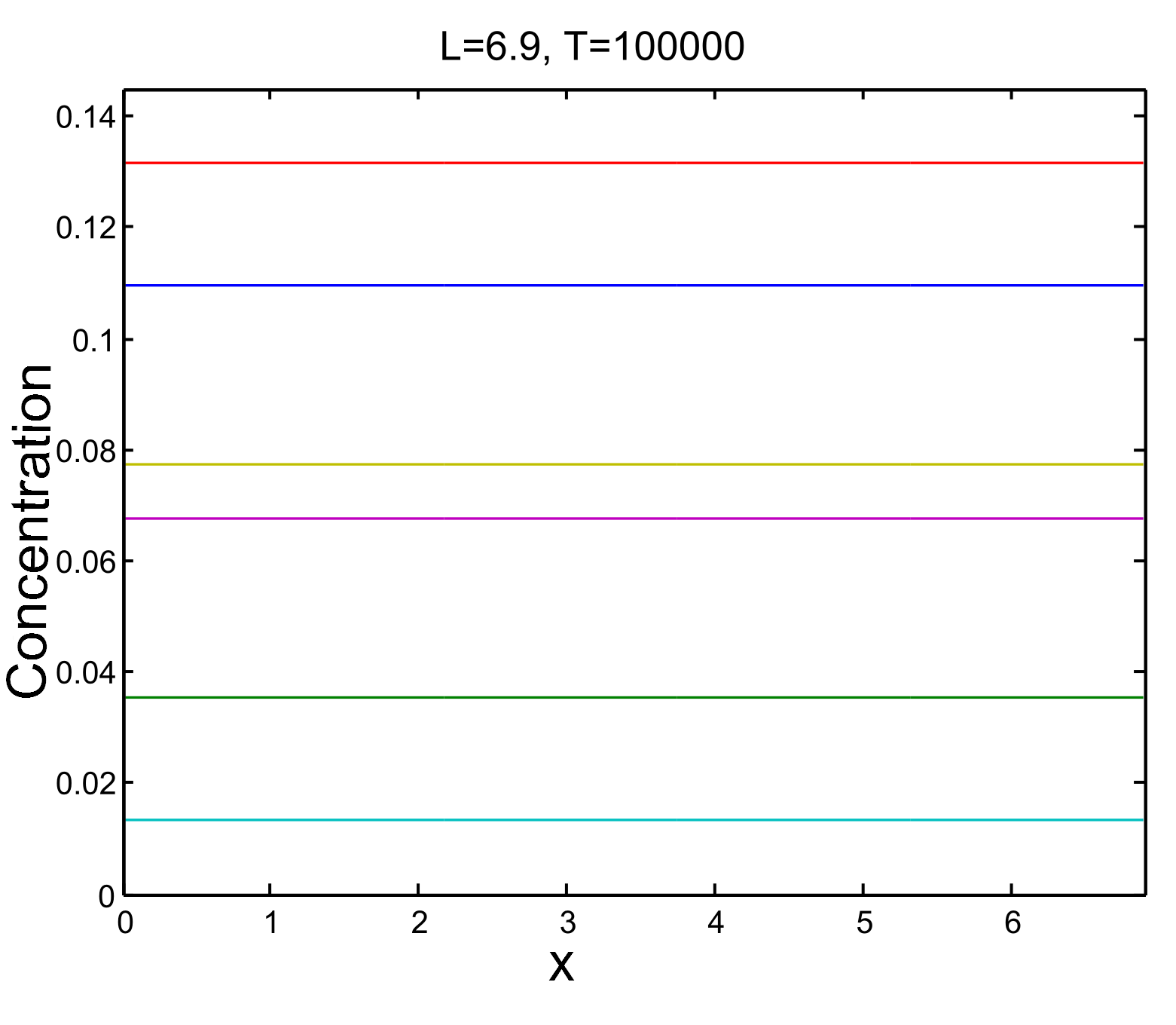}\includegraphics[width=0.3\textwidth]{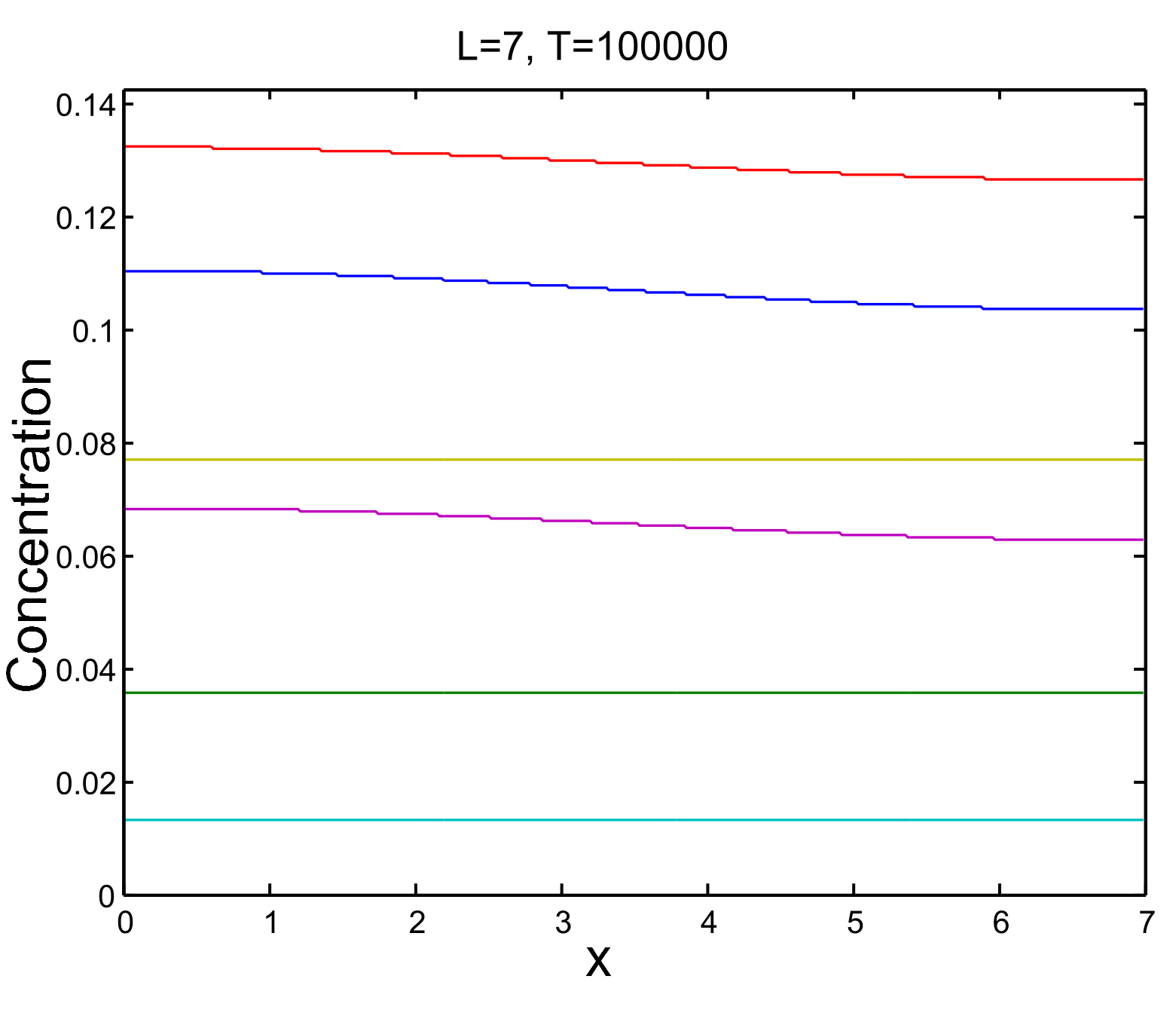}
\includegraphics[width=0.3\textwidth]{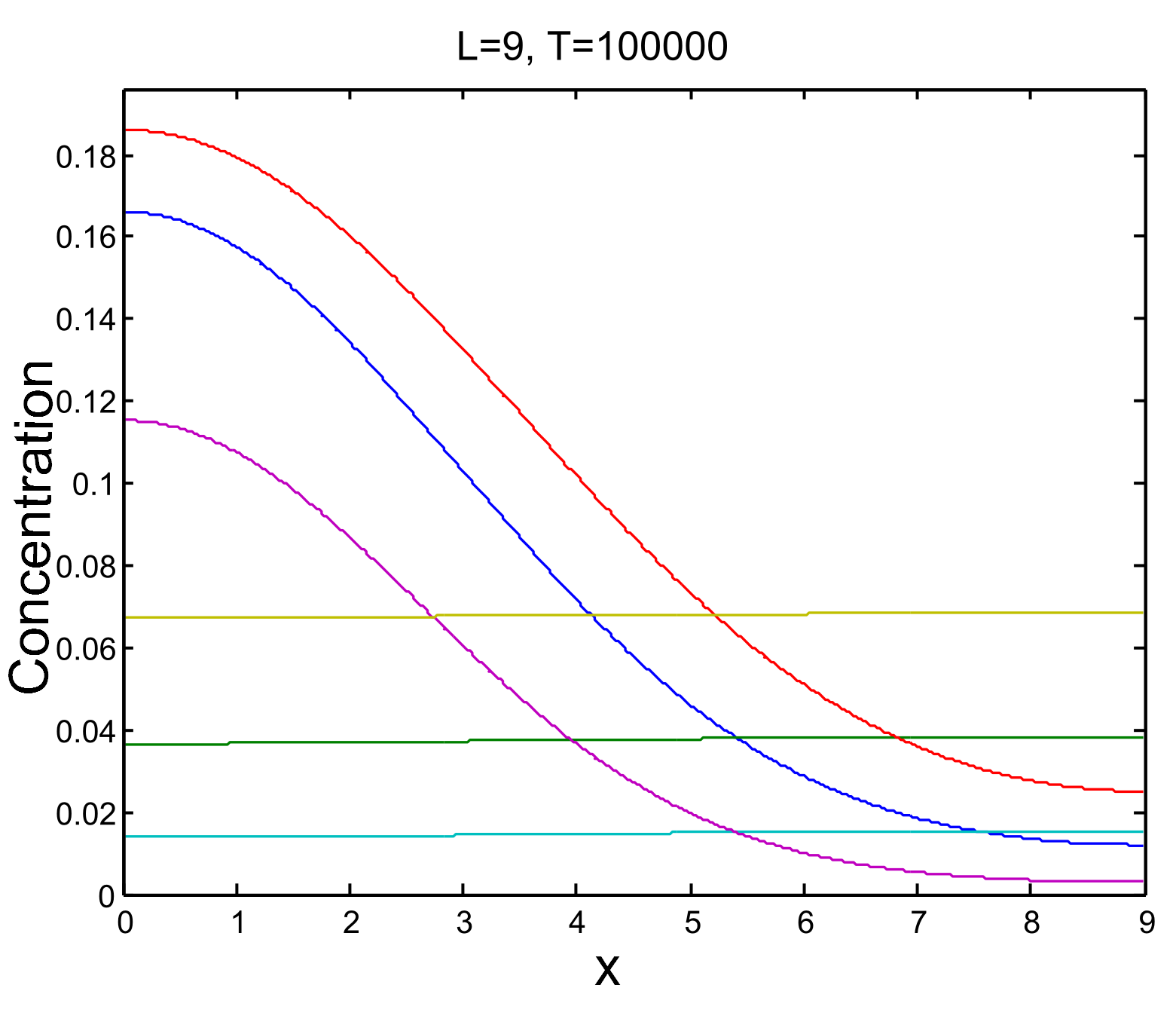}
\includegraphics[width=0.3\textwidth]{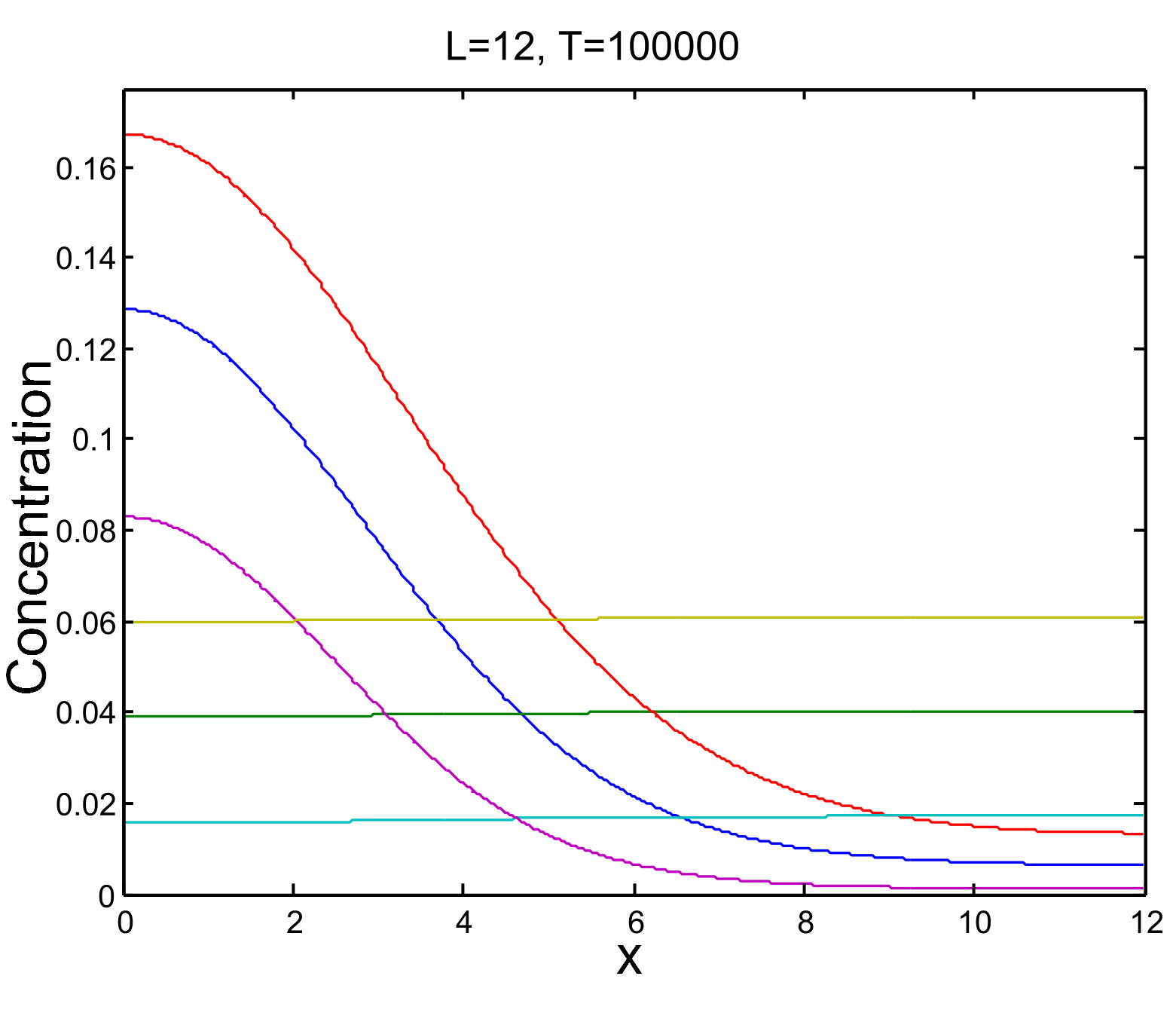}\includegraphics[width=0.3\textwidth]{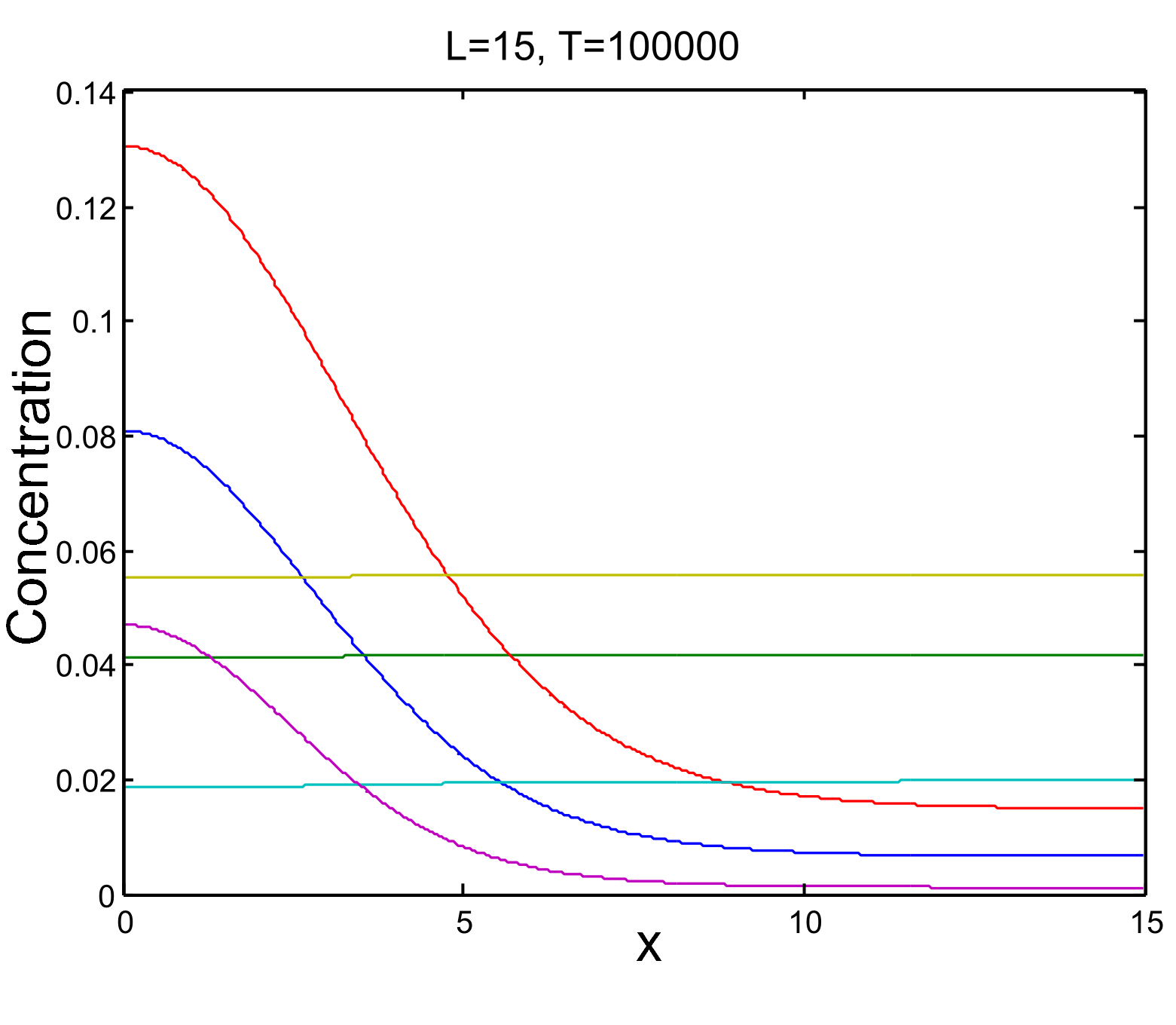}\caption{\label{HDPolarisierung}Stationary concentrations for different sizes of $L$.}
\end{center}
\end{figure}

In order to investigate the ability to reproduce length-dependent polarization we run the system with initial concentrations $u_i = 0.5 + 0.001 \cos \pi x$ for different sizes of $L$ until a stationary state is reached ($t=100000$ seems to be a reasonable choice). The results are given in Figure \ref{HDPolarisierung}. One observes that for the parameter set given in Table \ref{Tab:homPa} length-dependent polarization occurs at around $L=7$. For smaller values the constant solution is stable, while for larger values a peak close to the left boundary (due to the higher perturbations on the left) appears, which is more and more pronounced for increasing $L$. To illustrate the time evolution leading to these stationary states we plot the concentrations at several intermediate time steps for the case $L=12$. 

\begin{figure}[h]
\begin{center}
\includegraphics[width=0.3\textwidth]{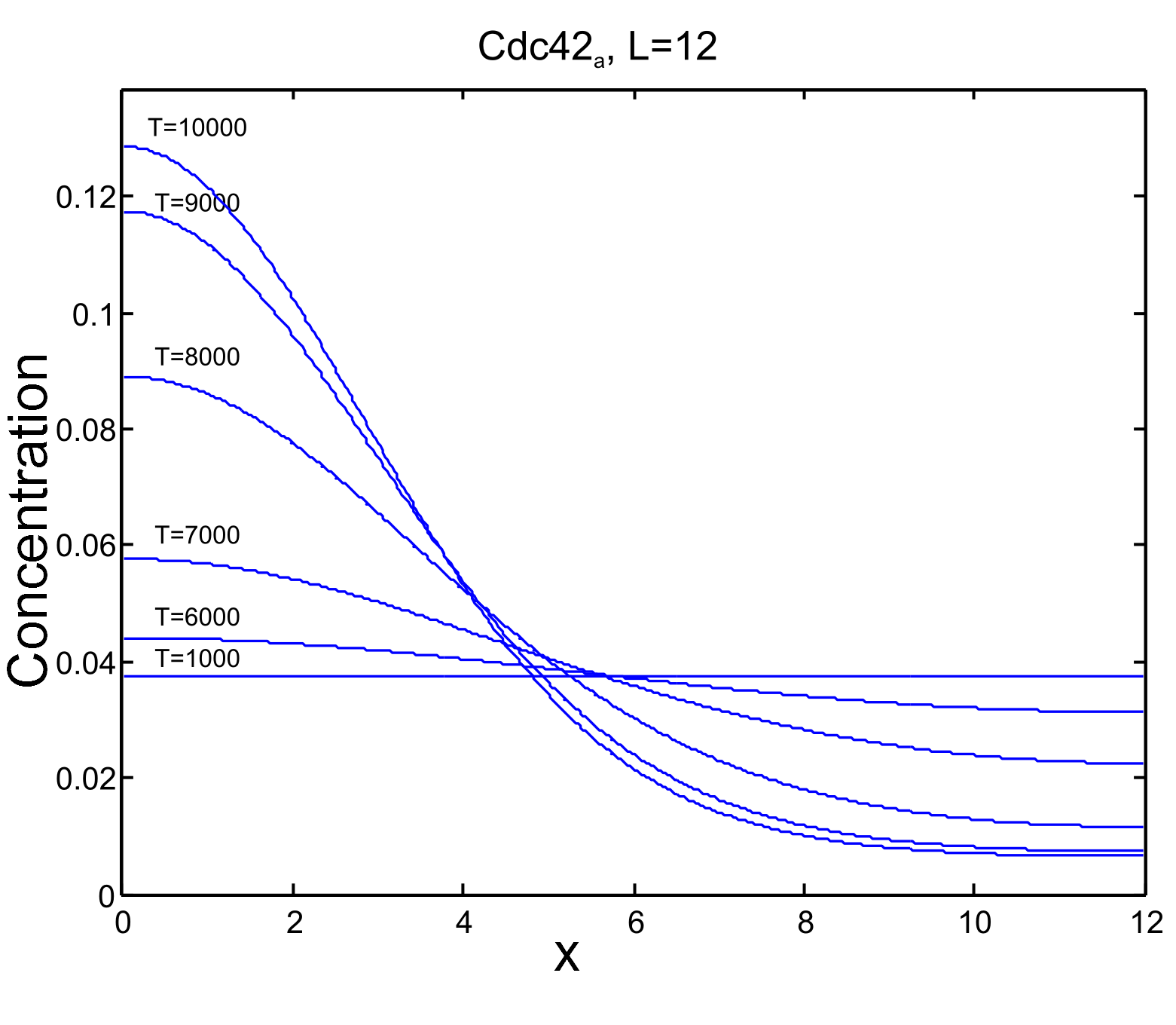}
\includegraphics[width=0.3\textwidth]{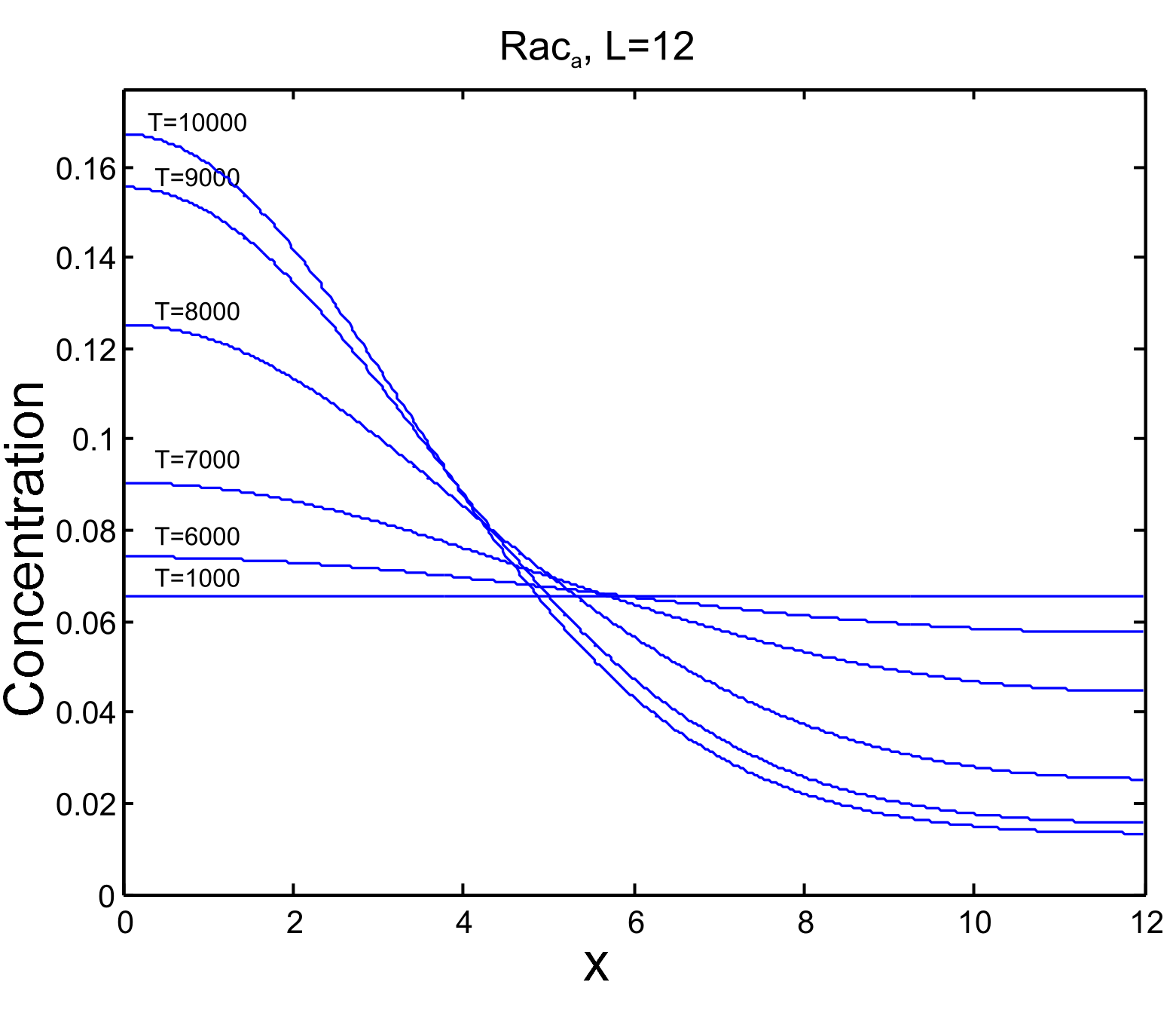}\includegraphics[width=0.3\textwidth]{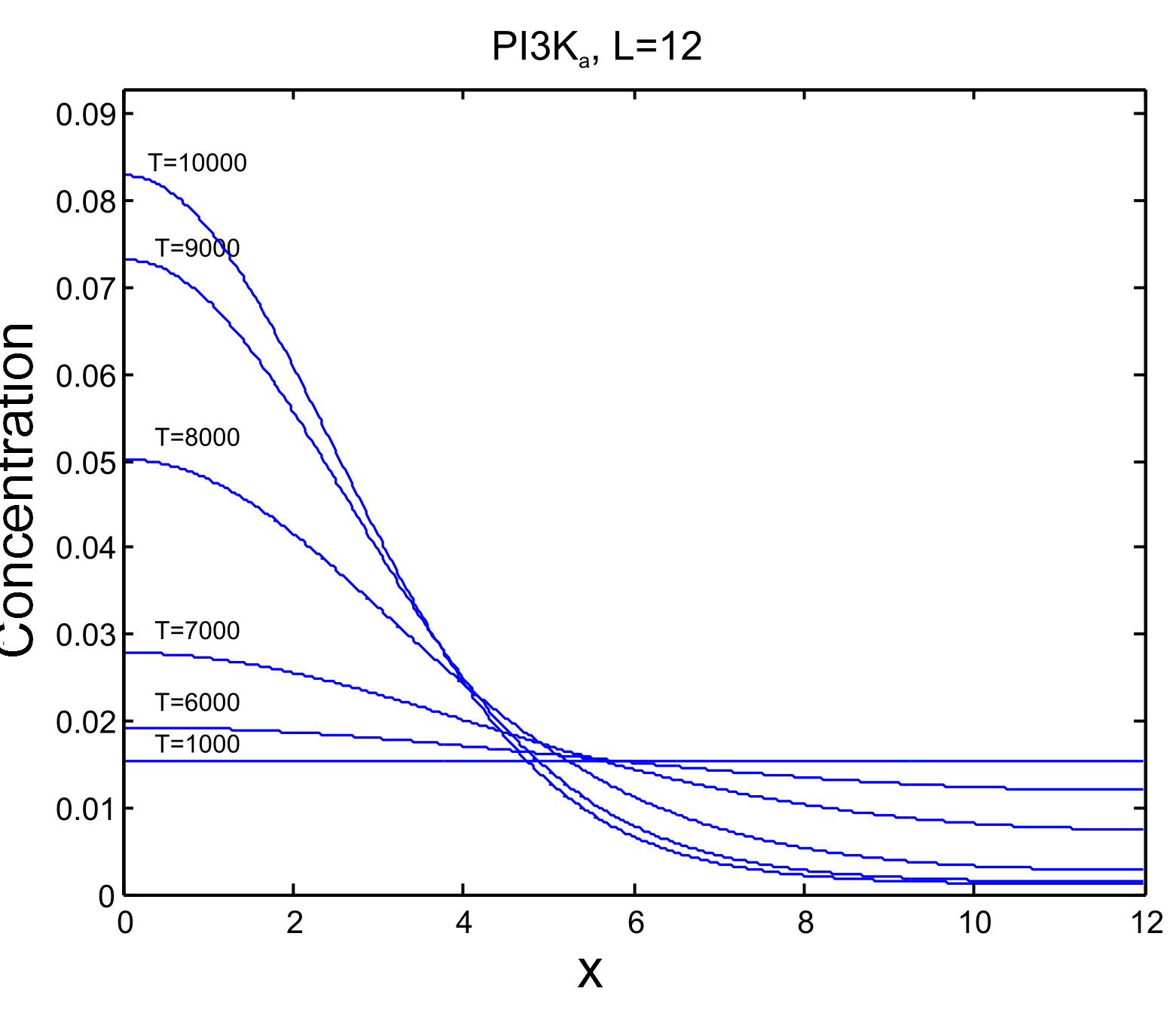}
\caption{\label{HDZeitschritte}Evolution of active Cdc42, Rac and PI3 kinase concentrations in a case leading to polarization.}
\end{center}
\end{figure}

\begin{figure}[h]
\begin{center}
\includegraphics[width=0.3\textwidth]{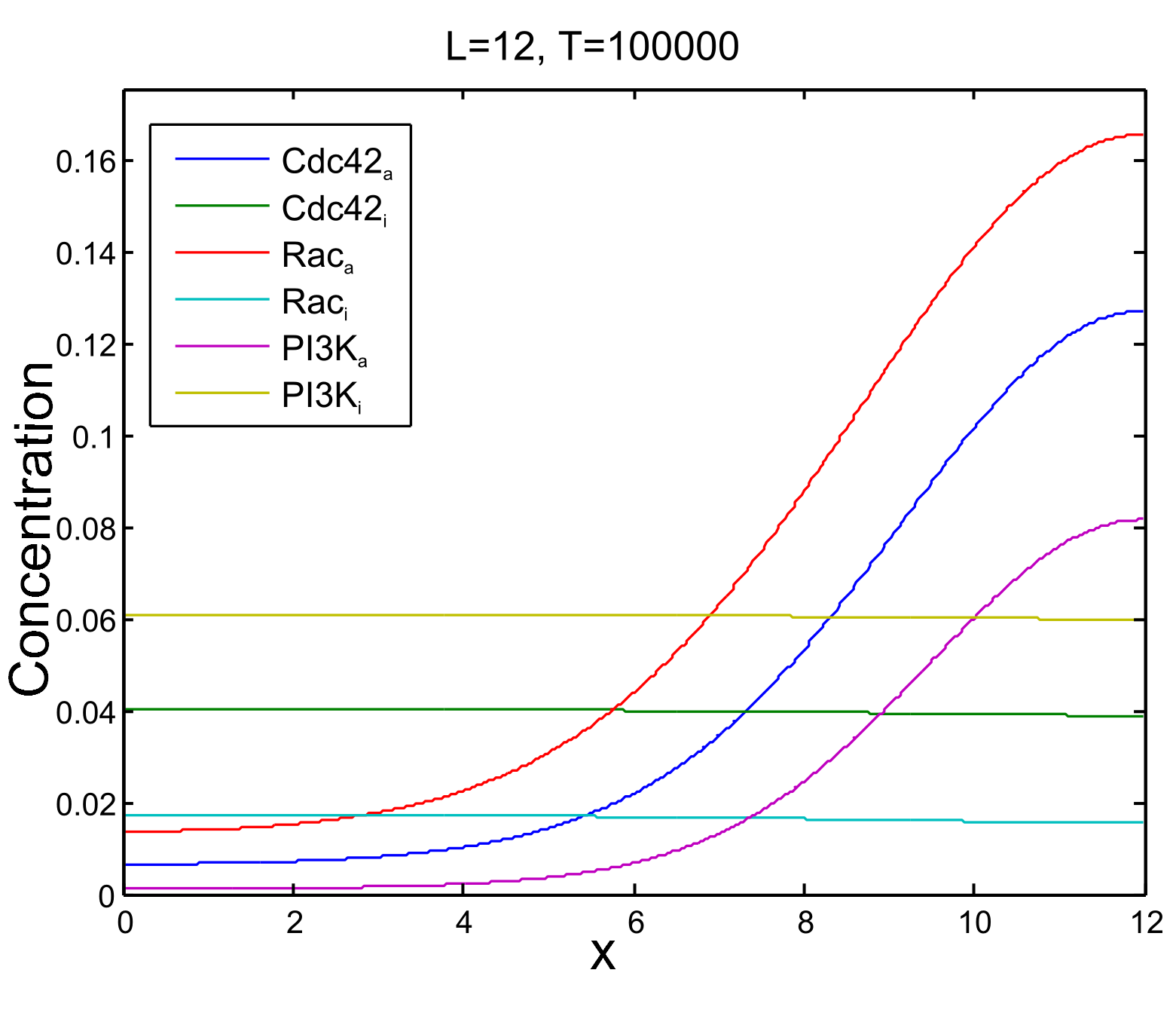}
\includegraphics[width=0.3\textwidth]{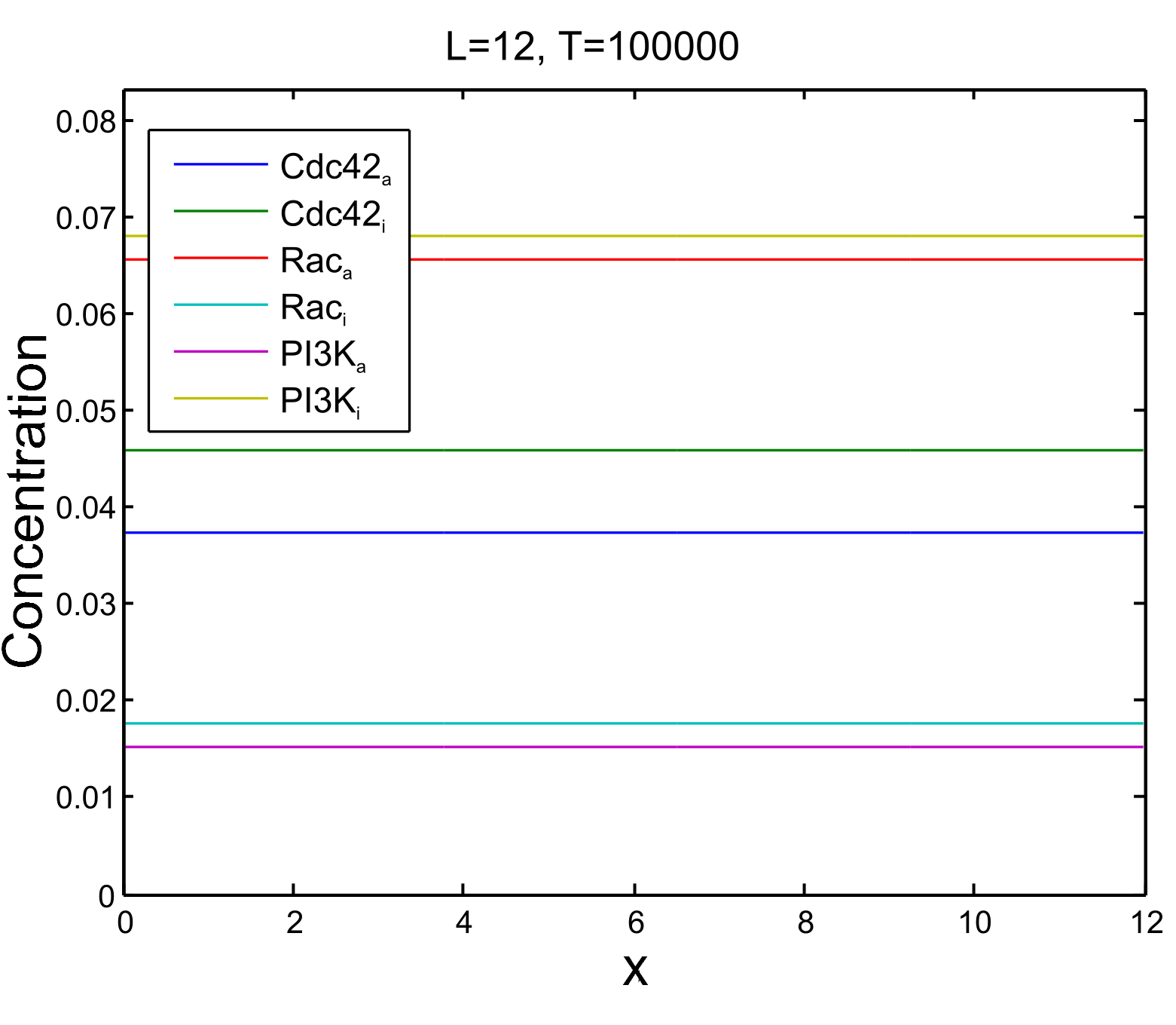}
\caption{\label{RauschenHochfrequent}Stationary concentrations for a (pointwise uniformly) random perturbation (left) and a high frequency perturbation (right).}
\end{center}
\end{figure}

Clearly the specific perturbation $\cos \pi x$ has a certain influence on the selection of the stationary solution, since e.g. for a perturbation with opposite sign one would end up with a solution mirrored at $x=\frac{1}2$ due to the inherent symmetry in the system. However, apart from the selection of a solution with left or right peaks, the perturbation seems not to be too crucial. To test this we use various random perturbations (uniformly distributed in each grid point) and the resulting stationary solutions are always the same as for the low-frequency cosine perturbation, except that with same probability we also obtain the mirrored one. To illustrate this we plot the resulting stationary concentrations for $L=12$ in the left panel of Figure \ref{RauschenHochfrequent}. The only type of perturbation that does not yield peak solutions also for large $L$ is a deterministic high-frequency one. This is not surprising since for large pure frequencies the diffusion dominates and smoothes the perturbation. This feature is illustrated in the right panel of Figure \ref{RauschenHochfrequent} with a plot of the resulting stationary concentrations in the case $L=12$.

\section{Symmetry Breaking and Axon Polarization}

\begin{figure}[h]
\begin{center}
\includegraphics[width=\textwidth]{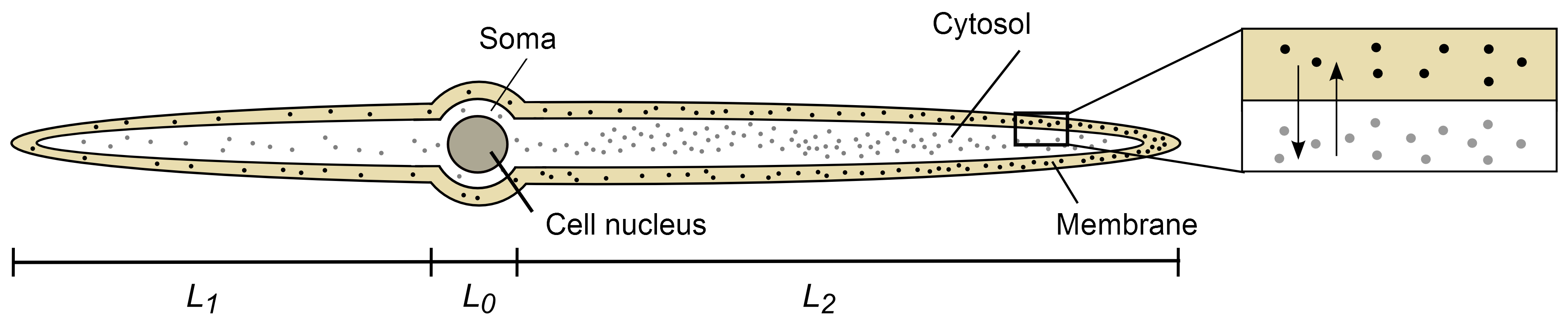}
\caption{\label{AxonModell}Setup of a neuronal cell with two developing neurons of lengths $L_1$ and $L_2$, with a soma of length $L_0$ inbetween.}
\end{center}
\end{figure}

In the previous section we have seen that the reaction diffusion model for PI3 kinase, Cdc42 and Rac is indeed able to reproduce length-dependent polarization, which is a first confirmative answer to the possible existence of an inherent mechanism. The remaining question is how the symmetry in the system is broken, in particular how the observation that polarization to a peak appears usually at the tip of the longest neurite (cf. \cite{Do88}) can be explained. For this sake we extend the model in the following to deal with a setup including neurites of different lengths. Since we want to keep the setting reasonably simple we use a system with two neurites separated by the soma as sketched in \eqref{AxonModell}, which still seems reasonable to be reduced to one spatial dimension and for which there is also some biological motivation as highlighted in \cite{Ca08}. 

\subsection{Modelling the Soma} 
 
Our main rationale for including the soma into our model is that the mobility (and hence the diffusion coefficient) is lowered. For the cytosolic form this is due to the cell nucleus, which prevents flow in a large fraction of the soma, and for the membrane-bound form this is due to the additional curvature of the membrane along the soma, which is otherwise not incorporated in our model.
Consequently we change all terms $\partial_{xx} u_i$ in (\ref{eqn:Sys2a})-(\ref{eqn:Sys2b}) to $\partial_x (d_i(x) \partial_x u_i)$, with 
\begin{equation}
d_i(x)=
\left\{
\begin{aligned}\label{Diffusionsparameter1}
1& \qquad \text{if} \quad 0<x<l_1 \quad \text{and}\quad (l_1+l_0)<x<1,\\
c_i < 1& \qquad \text{if} \quad l_1<x<(l_1+l_0).
\end{aligned}
\right.
\end{equation}

For illustrative numerical tests we choose the values $c_i=0.6$ for $i$ odd (membrane) and $c_i=0.3$ (cytosol) and all other parameters as in the previous section, but similar results have been found for other settings. For more intuitive visualization we rescale the lengths on the $x$-axis in all plots and center the system such that the mid point of the soma is located at $x=0$, in addition we use a light coloring of the soma region. 
We again use initial values $u_i(x,0)=0.5 + 0.001 \cos \pi x$.

\begin{figure}[h]
\begin{center}
\includegraphics[width=0.3\textwidth]{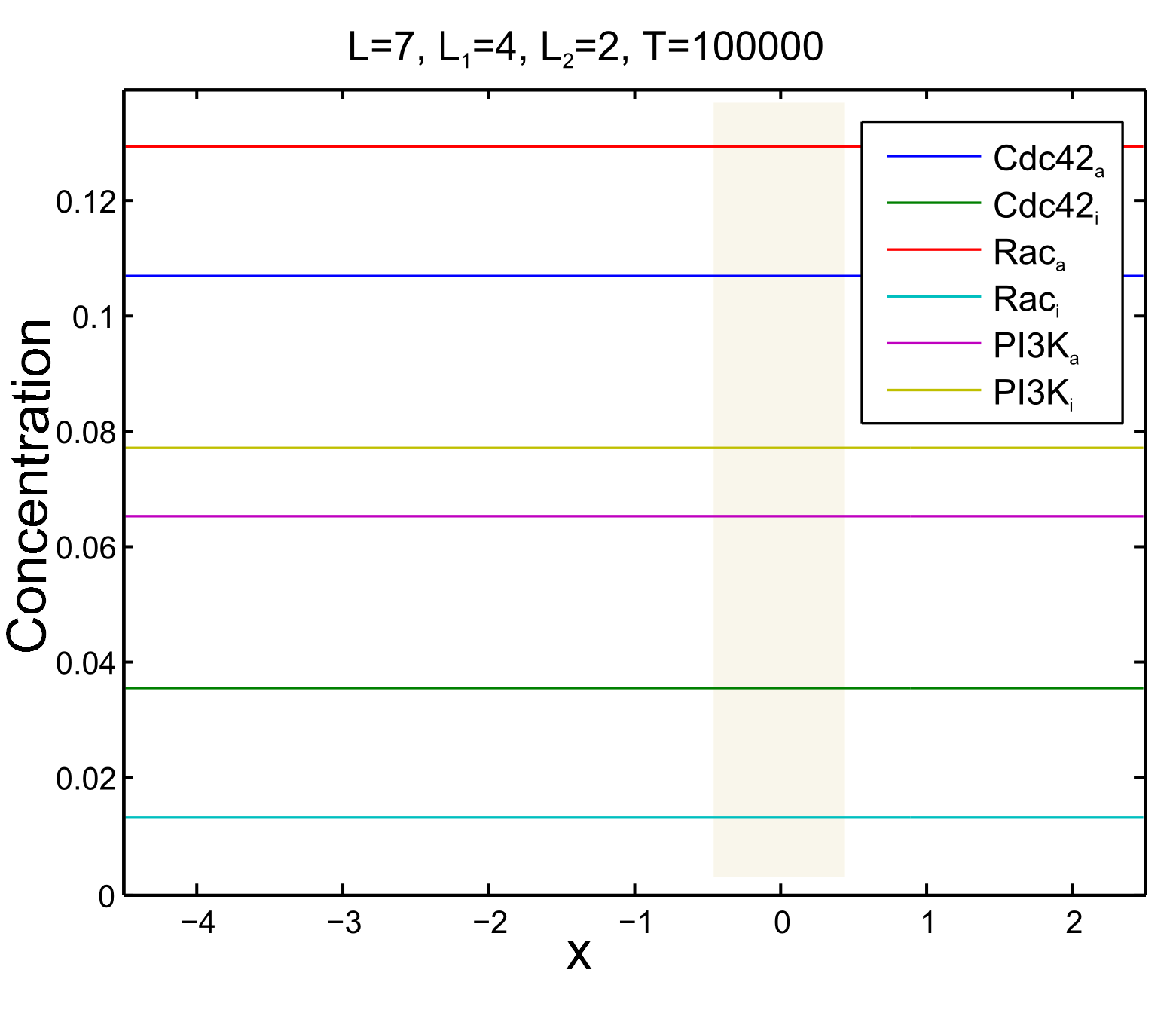}
\includegraphics[width=0.3\textwidth]{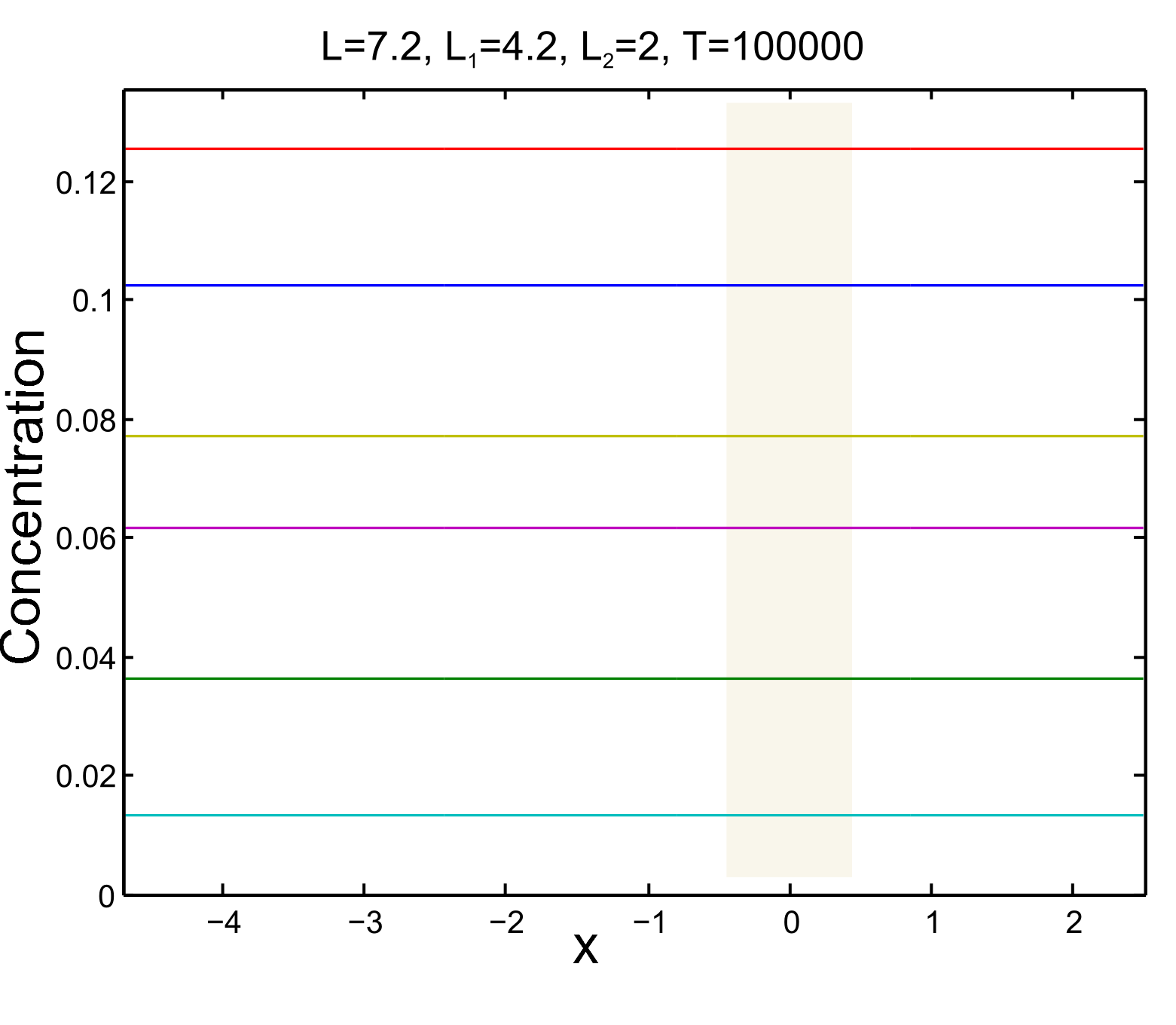}\includegraphics[width=0.3\textwidth]{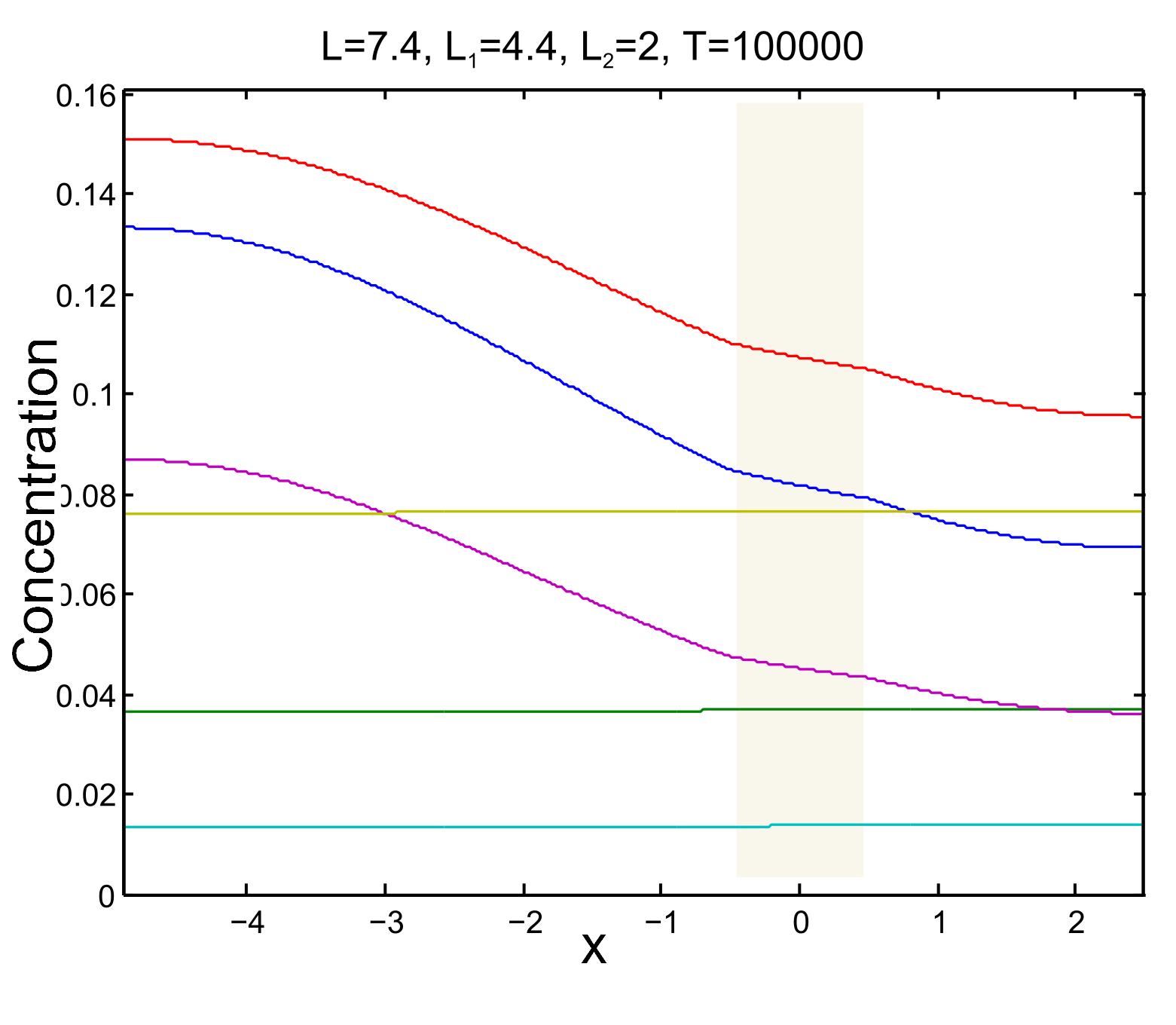}
\includegraphics[width=0.3\textwidth]{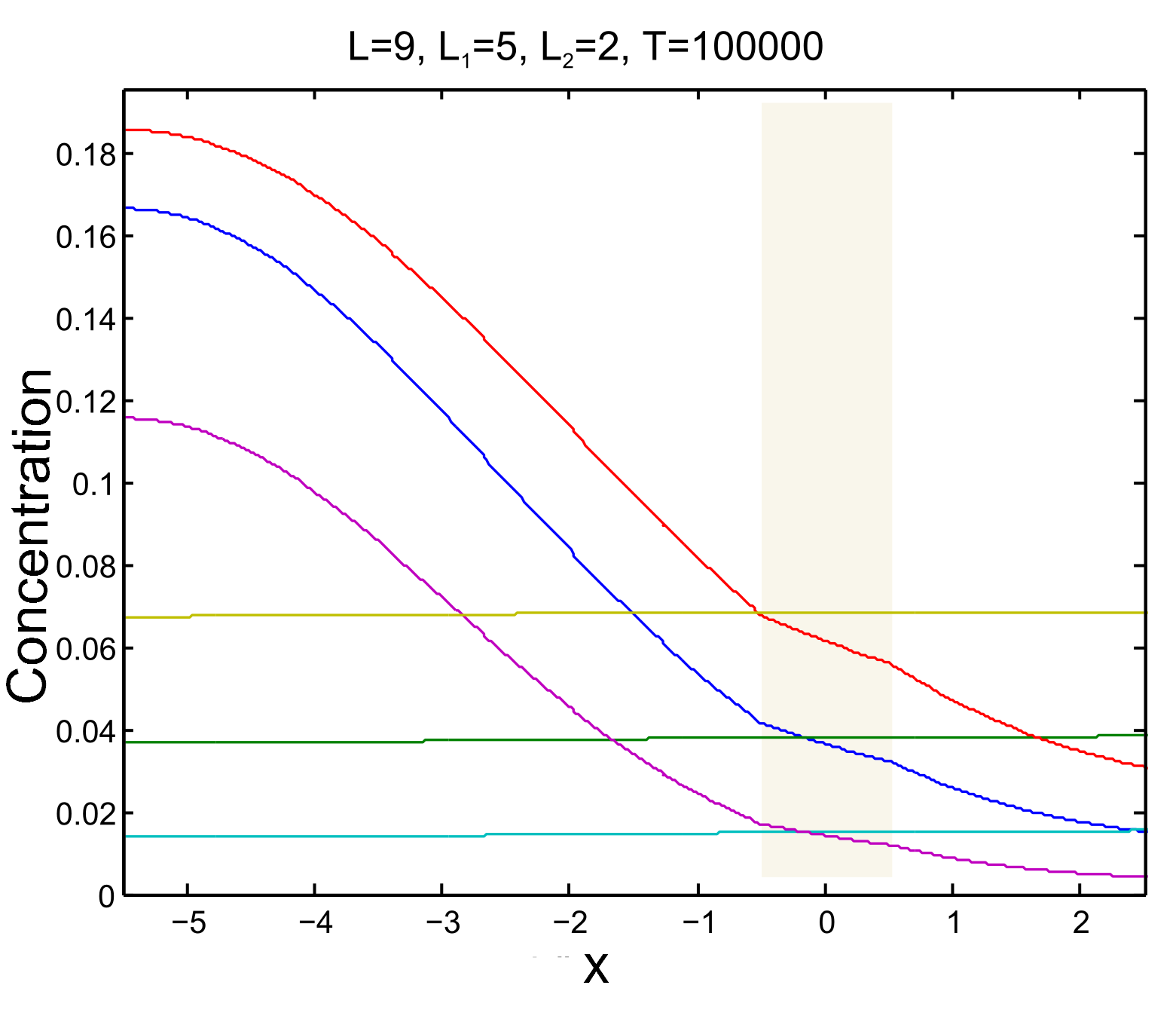}
\includegraphics[width=0.3\textwidth]{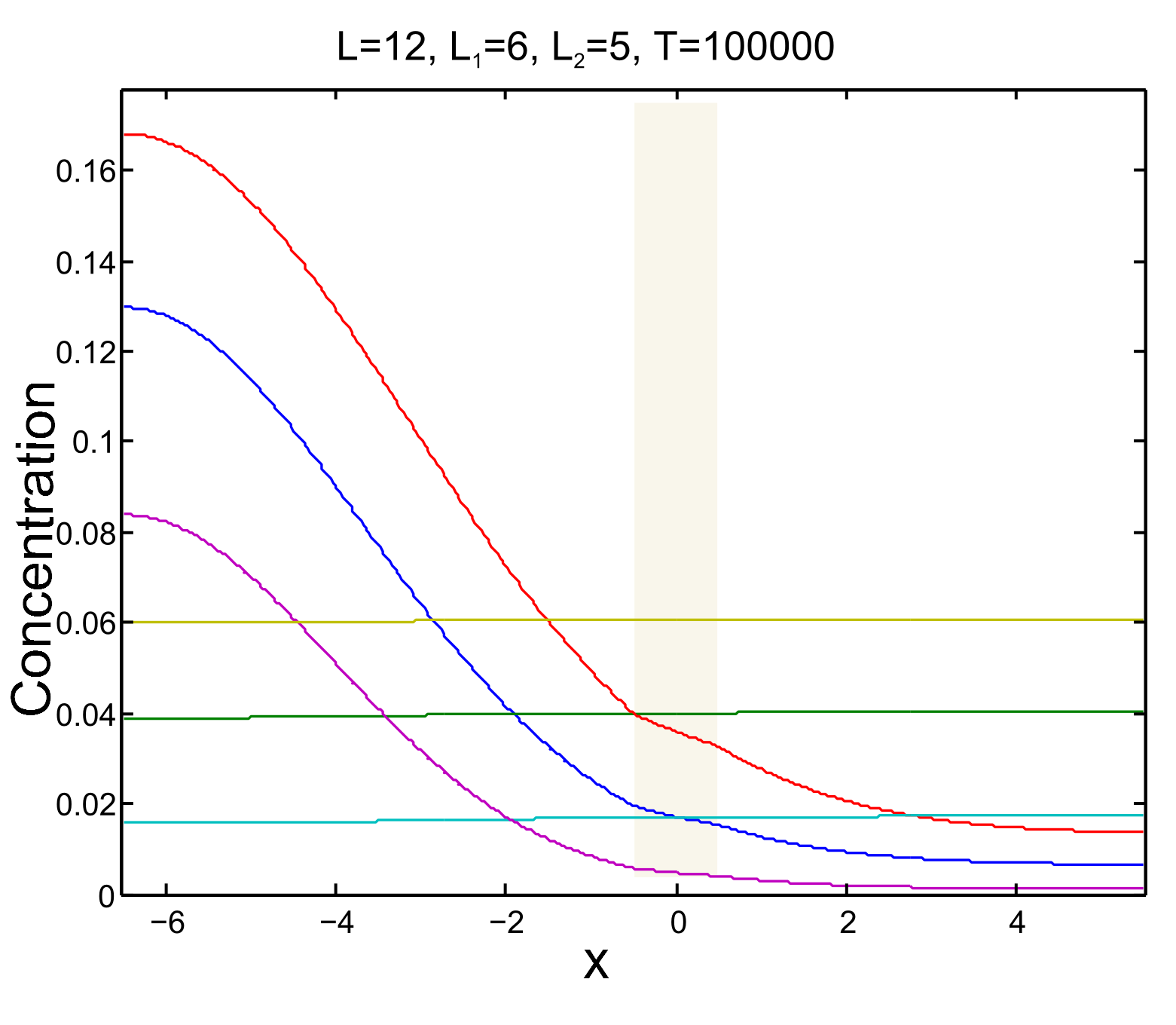}\includegraphics[width=0.3\textwidth]{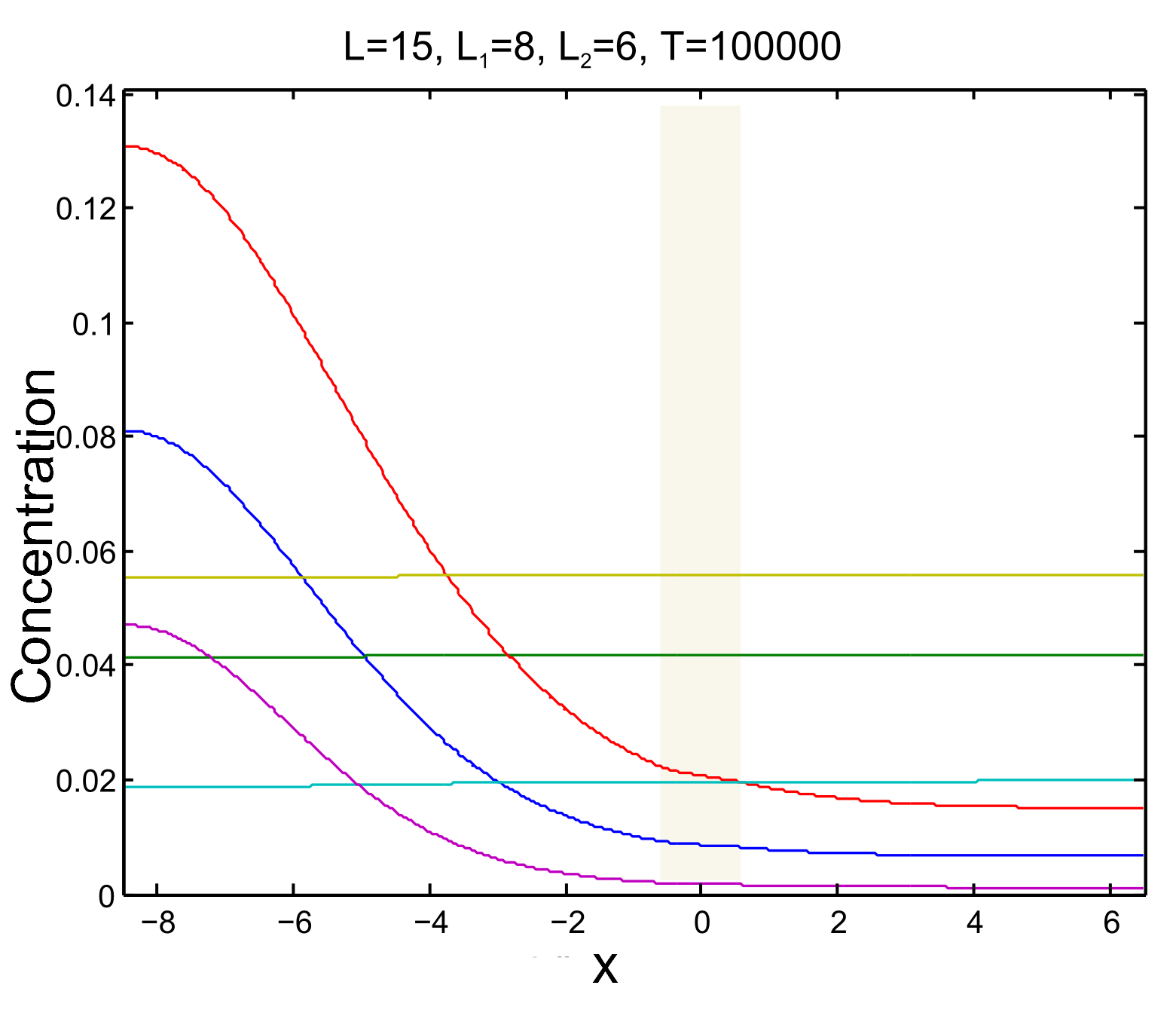}\caption{\label{IDPolarisierung}Stationary concentrations for different sizes of $L$.}
\end{center}
\end{figure}

\begin{figure}[h]
\begin{center}
\includegraphics[width=0.3\textwidth]{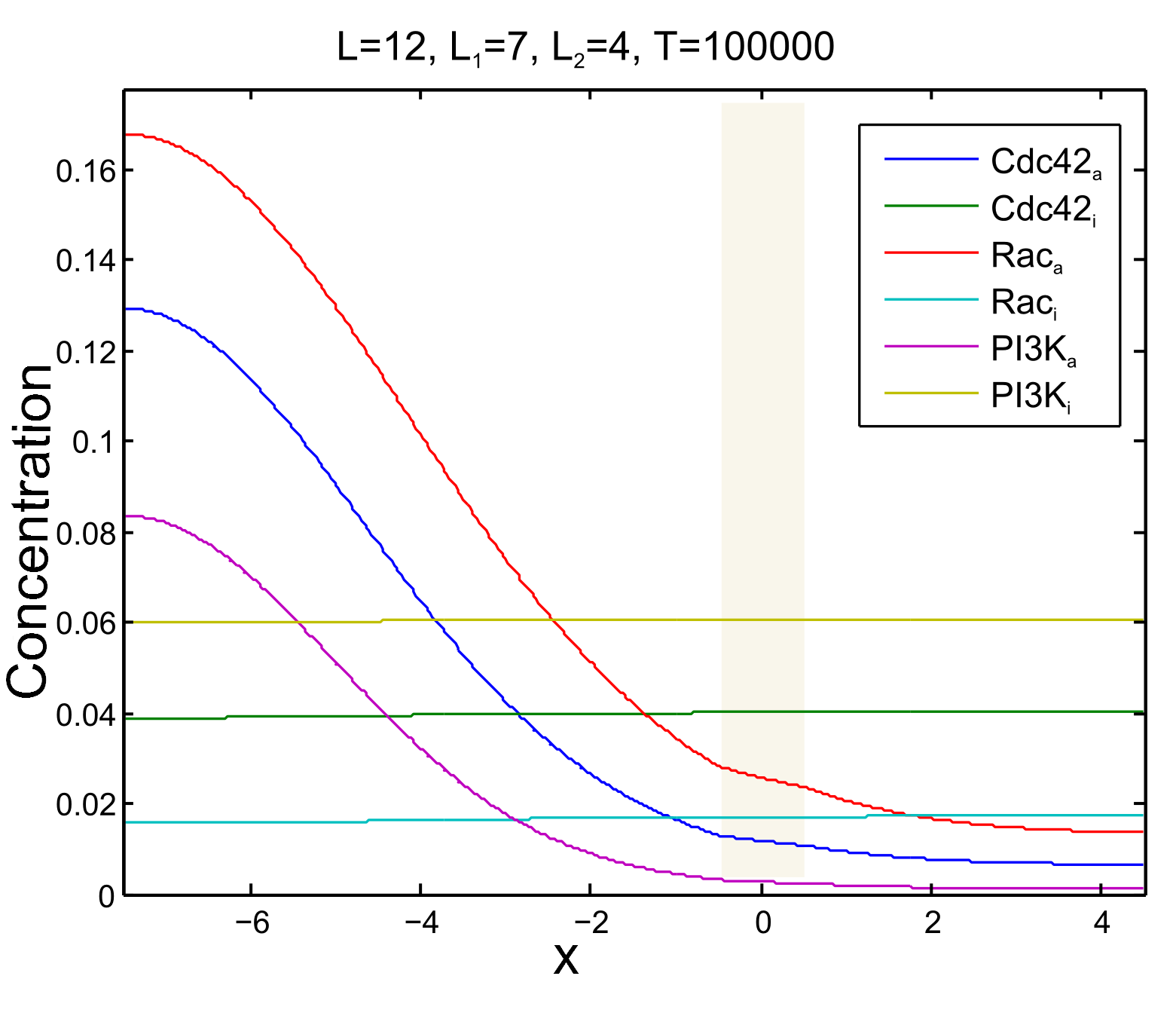}
\includegraphics[width=0.3\textwidth]{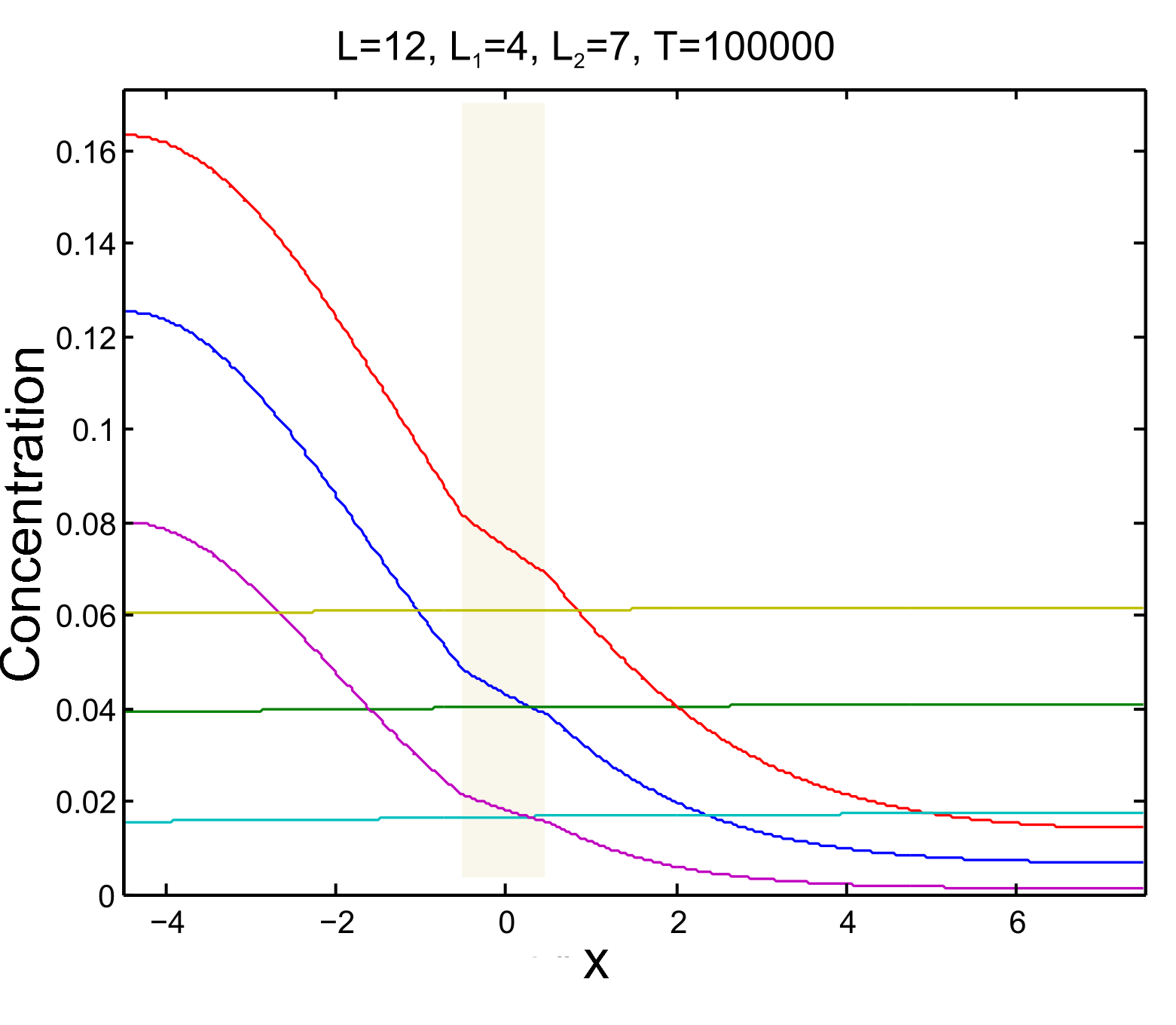}\includegraphics[width=0.3\textwidth]{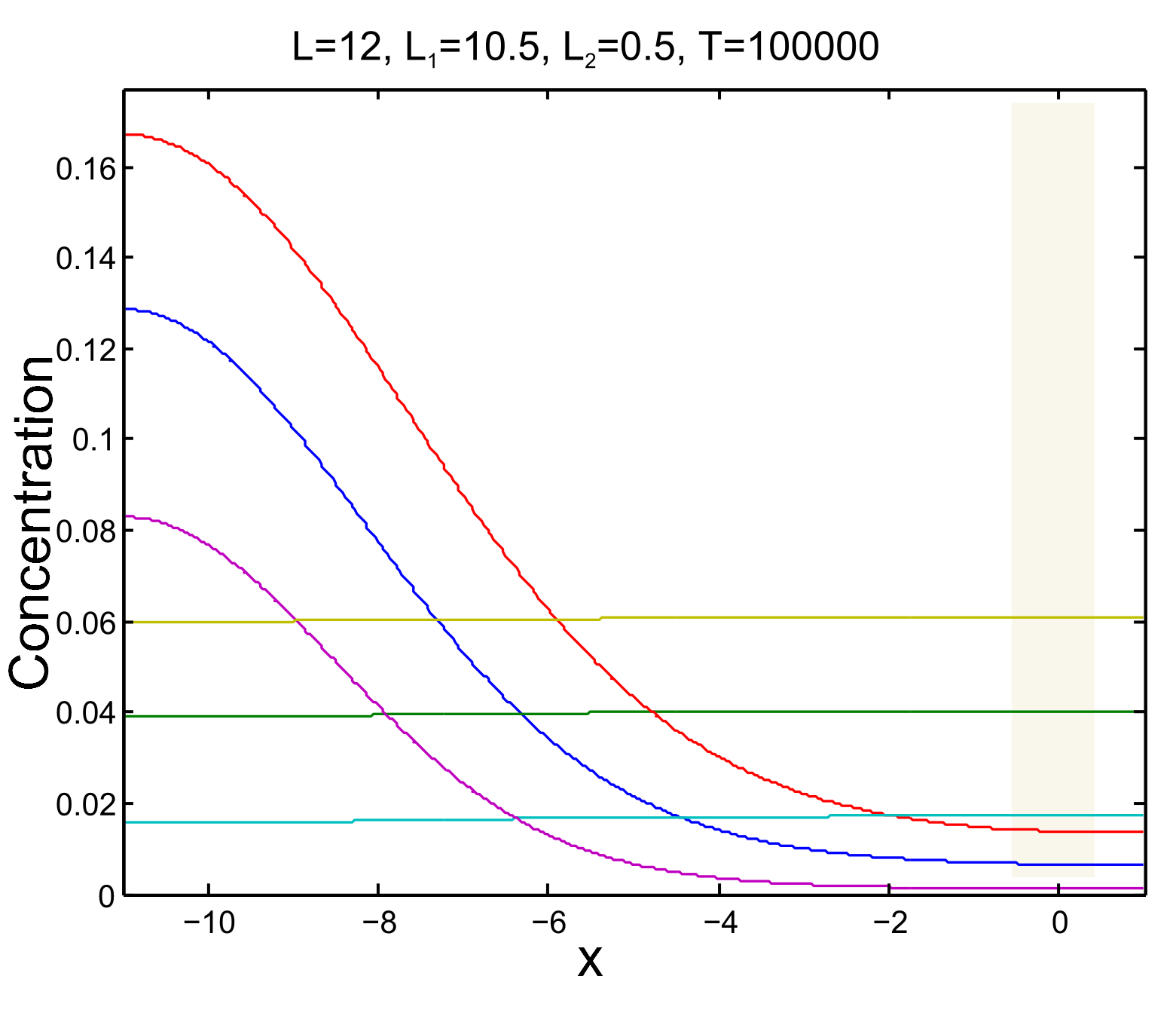}
\includegraphics[width=0.3\textwidth]{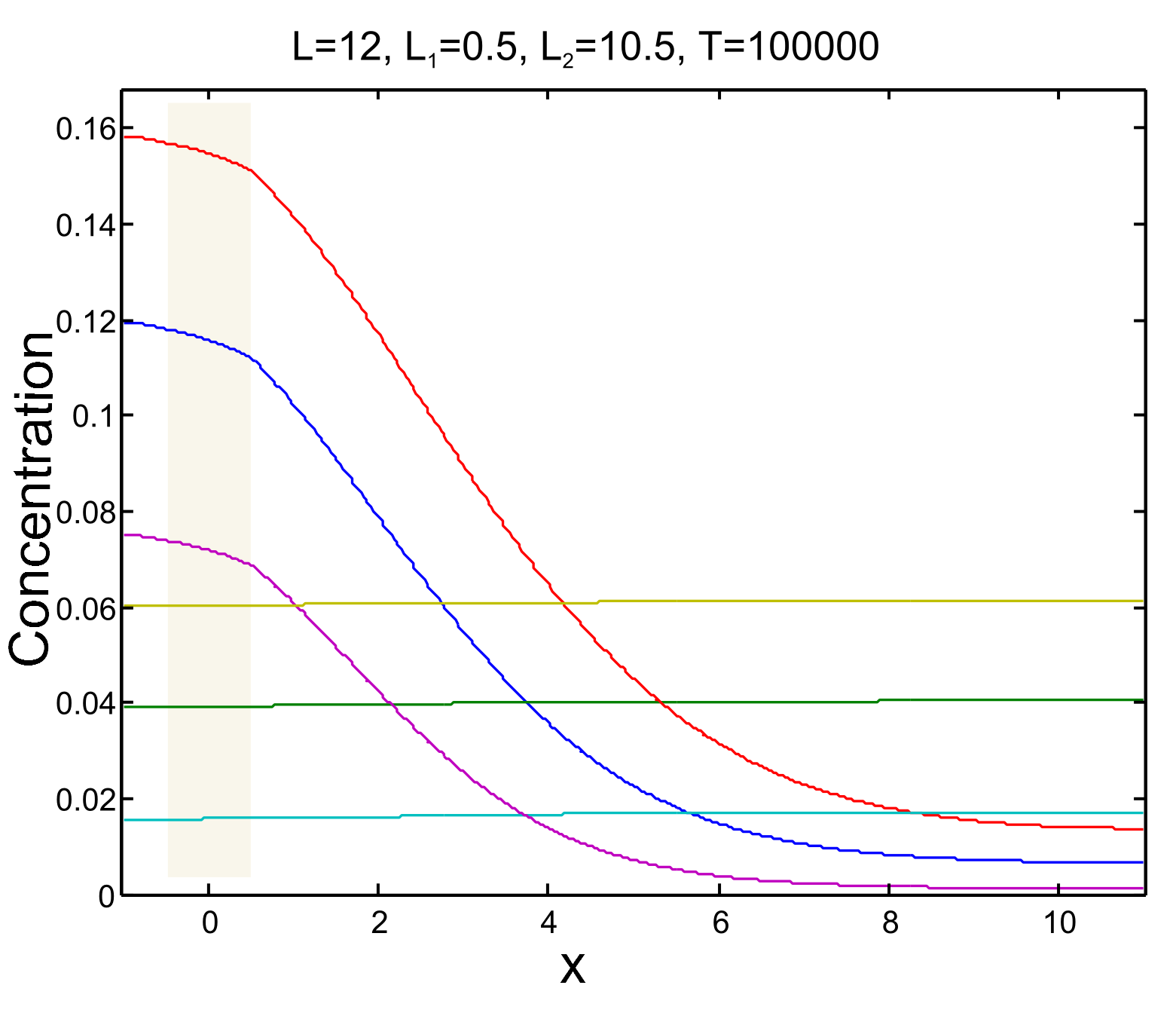}
\includegraphics[width=0.3\textwidth]{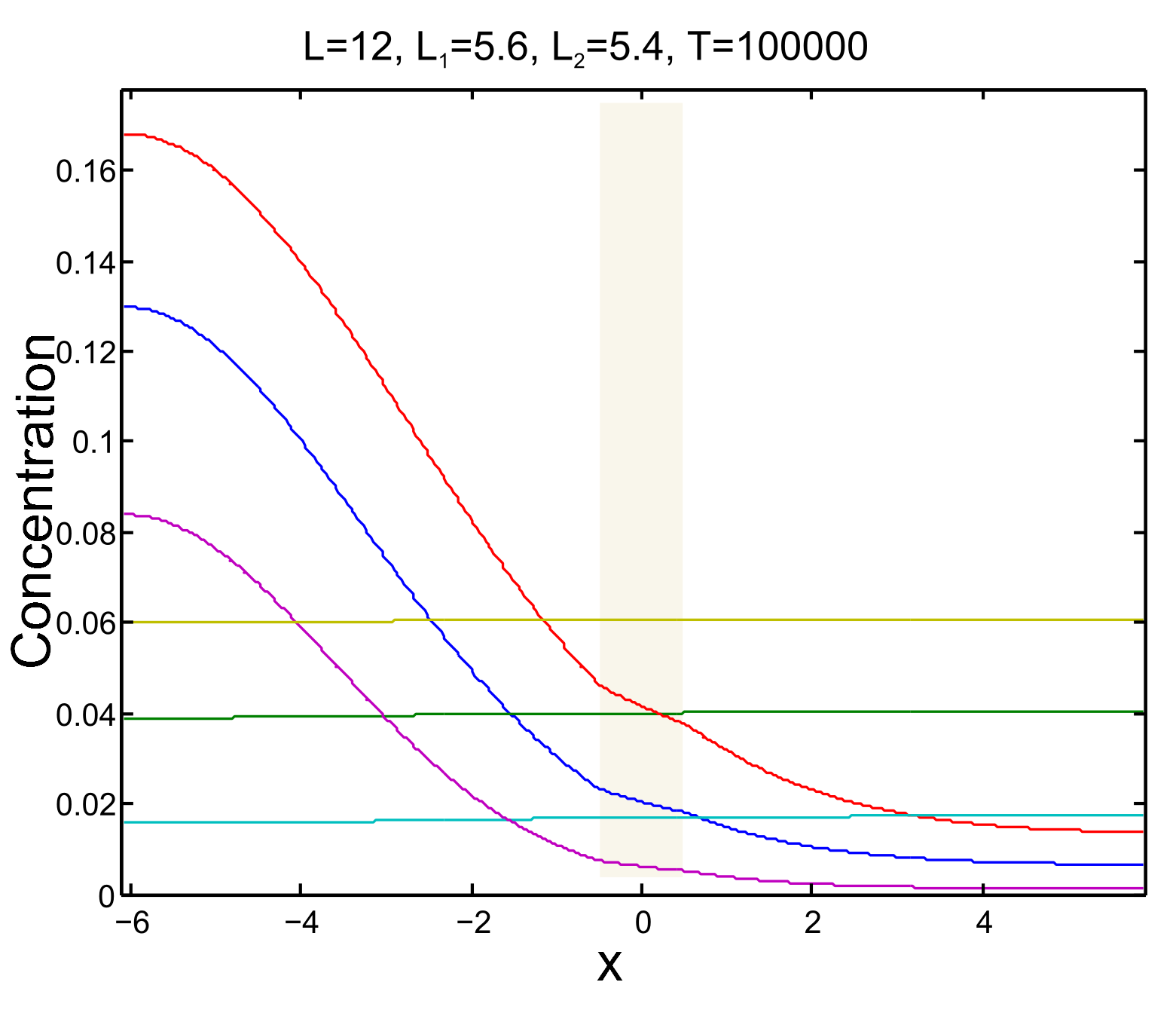}\includegraphics[width=0.3\textwidth]{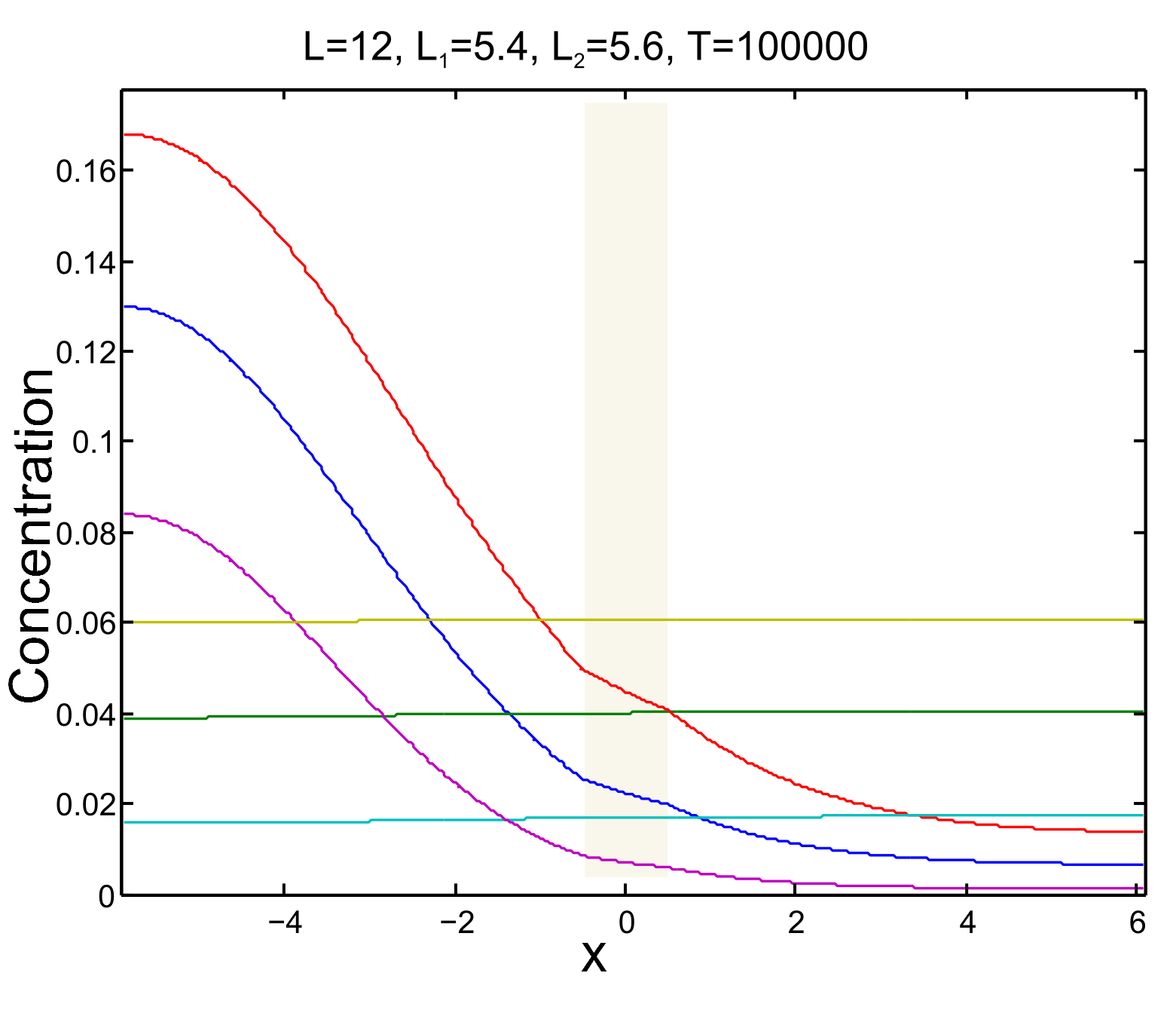}\caption{\label{IDPolarisierung2}Stationary concentrations for constant length $L$ and varying neurite lengths $L_1$ and $L_2$.}
\end{center}
\end{figure}

Figure \ref{IDPolarisierung} demonstrates that still length-dependent polarization, in this case at around a length $L=7.3$ of the overall system, appears. The increase in the critical length compared to Figure \ref{HDPolarisierung} is due to the fact that the overall diffusion is smaller due to the soma. In these examples polarization occurs in the (longer) left neurite. However, there is no symmetry breaking towards polarization in the longer neurite, but the effect is caused by the positive initial perturbation in the left part. In a variety of simulations with different perturbations and length distributions we observed similar behaviour as in the model without soma, in particular a nonsymmetric initial perturbation most strongly influences the choice of the polarization. This is illustrated in Figure \ref{IDPolarisierung2} with results varying the lengths $L_1$ and $L_2$, while keeping all other parameters - in particular the shape of the perturbation - as above. 
A conclusion from these simulations is that the existence of a soma is not sufficient to obtain a symmetry-breaking towards polarization in the longer neurite as observed in experiments. We will thus consider a further extension in the next section.

\subsection{Active Transport}

Besides the mechanisms included in the model so far, it has been suggested that active anterograde transport has an important role in neuronal polarization (cf. \cite{Sa96,To10}). We will thus extend the model to include such a transport term. Our detailed motivation is a report of active transport of Shootin1 in \cite{Me04}, which has been identified to be a protein related to the localization of PI3 kinase (cf. \cite{To06}). Thus, we modify the (scaled) equation for the active PI3 kinase  concentration to 
\begin{align}
	\partial_t u_5 =&-\alpha_{9}u_5+\epsilon^2 \alpha_{10} u_3^2 u_{6}+ \alpha_{11}u_6+\epsilon^2 {\partial_x}\left(d_5(x) {\partial_x u_5}\right)\\ \notag &-\epsilon {\partial_x}(v(x)u_5),
\end{align}
with
\begin{equation}\label{eqn:Transport}
v(x)=
\left\{
\begin{aligned}
-v  & \qquad \text{if } \quad 0<x<l_1,&\\
v & \qquad \text{if} \quad (l_1+l_0)<x<l_2,&\\
0& \qquad \text{else.}&
\end{aligned}
\right.
\end{equation}
Since we are not aware of another experimental proof of active transport we keep the equations for the other concentrations unchanged here, but remark that our computational tests show similar results if active transport is added to any other equation. We also mention that here we use equal constant velocities in both neurites for simplicity. In \cite{To10} it was even suspected that the transport velocity for Shootin 1 is growing with the neurite length, which also does not change our qualitative results. 

\begin{figure}[h]
\begin{center}
\includegraphics[width=0.3\textwidth]{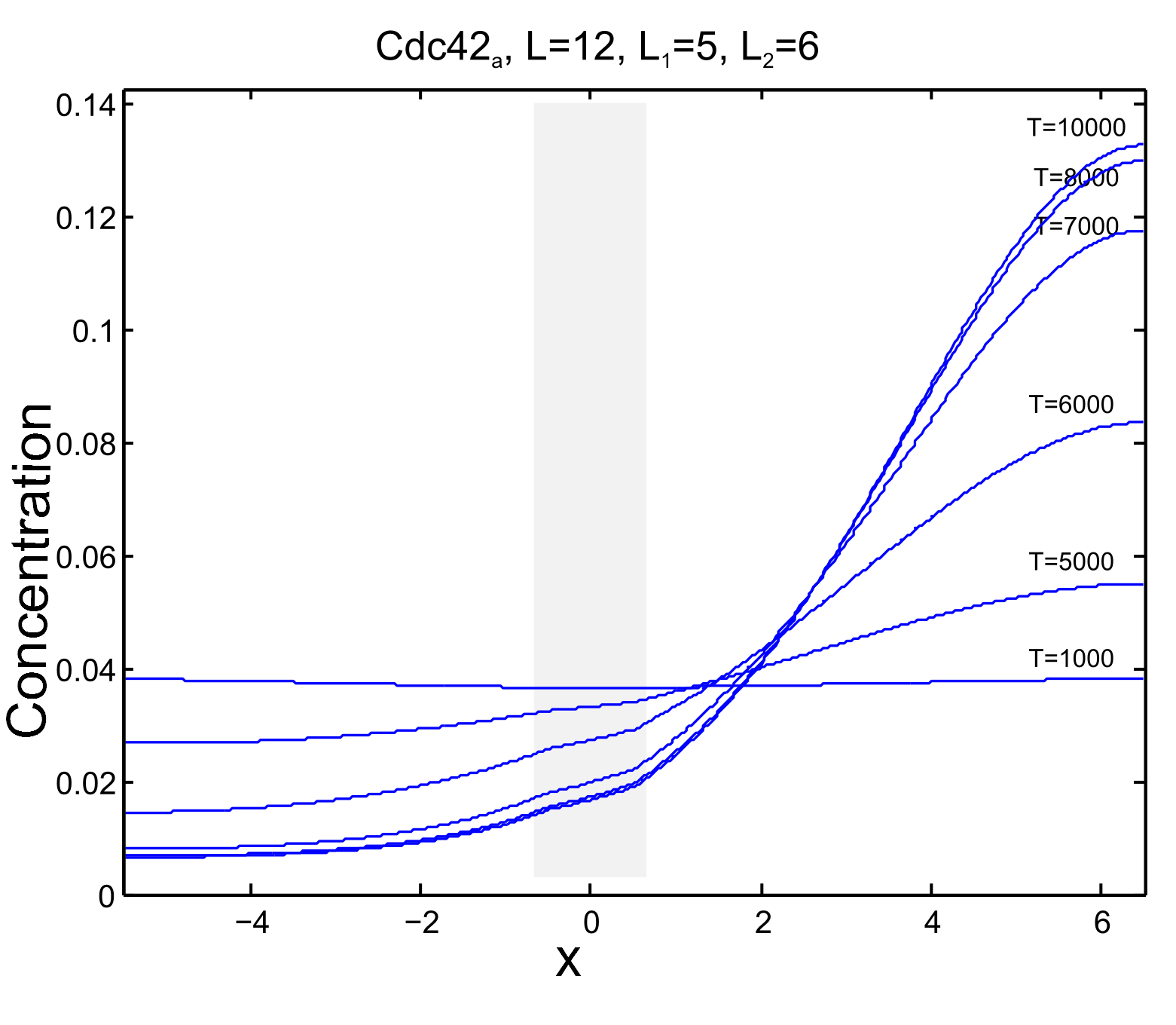}
\includegraphics[width=0.3\textwidth]{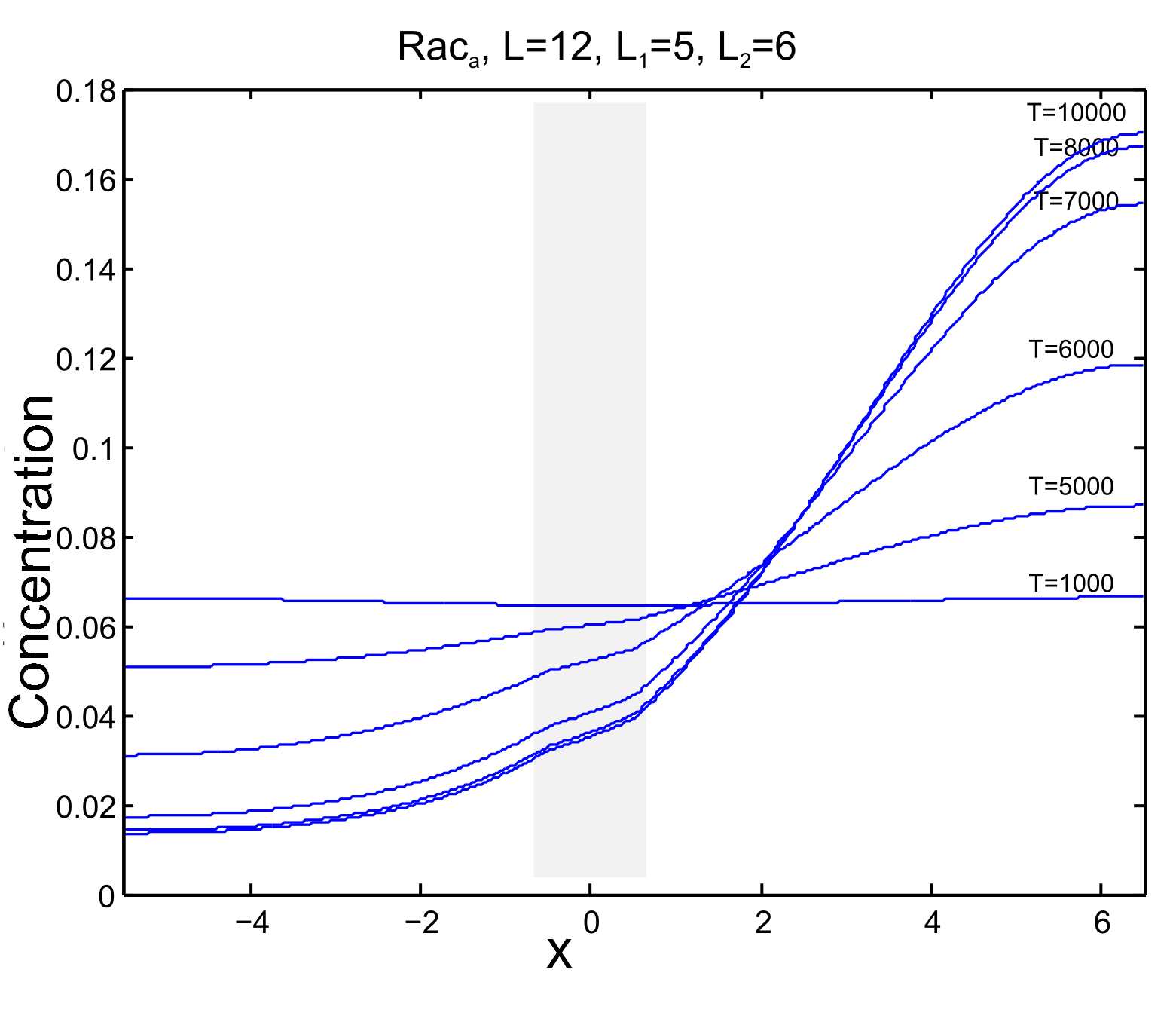}\includegraphics[width=0.3\textwidth]{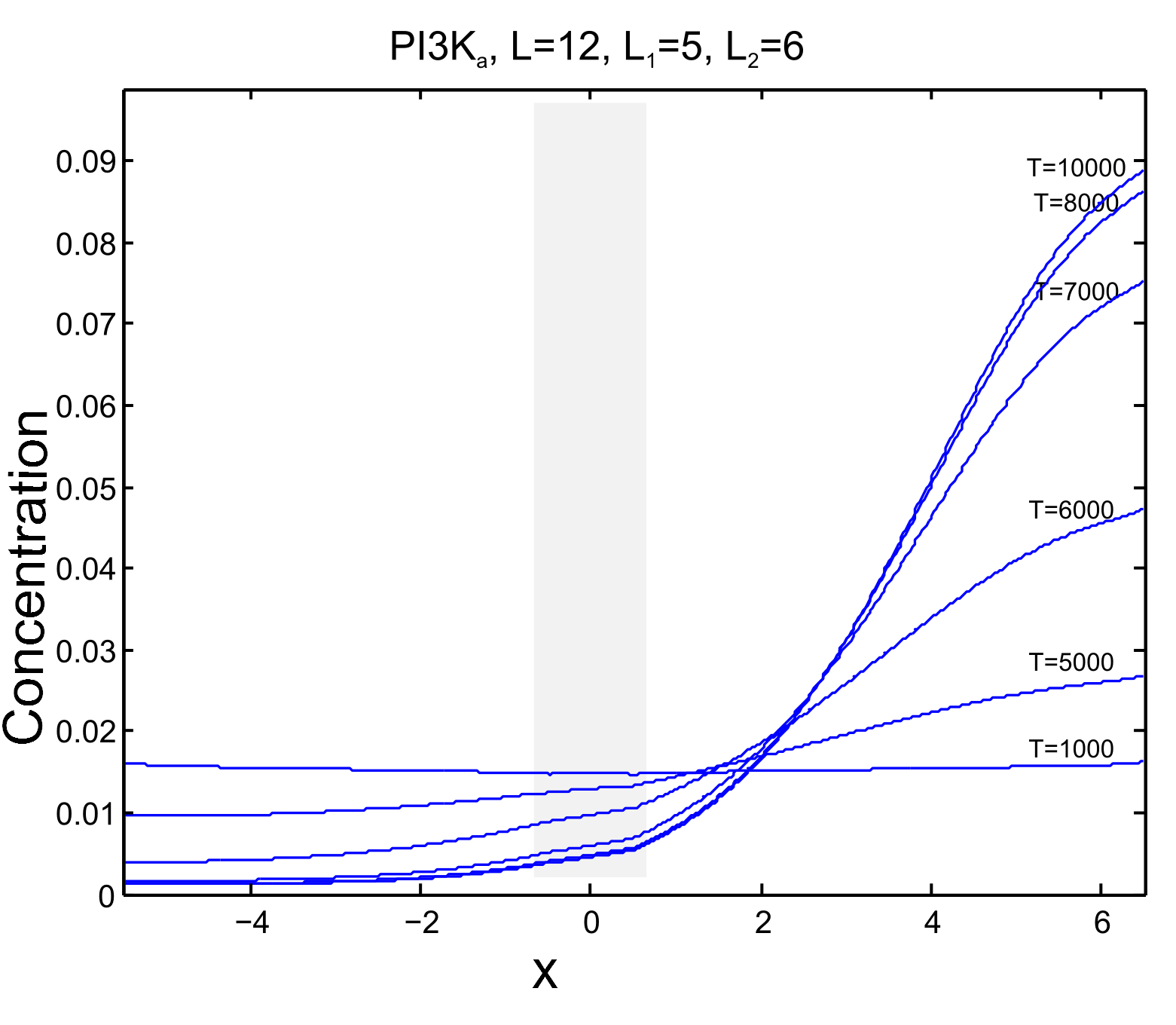}
\caption{\label{ATZeitschritte}Evolution of active Cdc42, Rac and PI3 kinase concentrations in a case leading to polarization.}
\end{center}
\end{figure}

The surprising result of using (even small) active transport is that symmetry breaking appears as in experiments and all observed effects are reproduced. The polarization now occurs mainly in the longer neurite even for positive perturbations in the shorter one. This is illustrated by the time evolutions of the active concentrations in Figure \ref{ATZeitschritte}, again with initial value $u_i(x,0) = 0.5 + 0.001 \cos \pi x$. The positive initial perturbation in the shorter left neurite reverted towards a positive perturbation in the longer neurite and subsequently a similar evolution towards the stationary state as in the symmetric case. 

\begin{figure}[h]
\begin{center}
\includegraphics[width=0.3\textwidth]{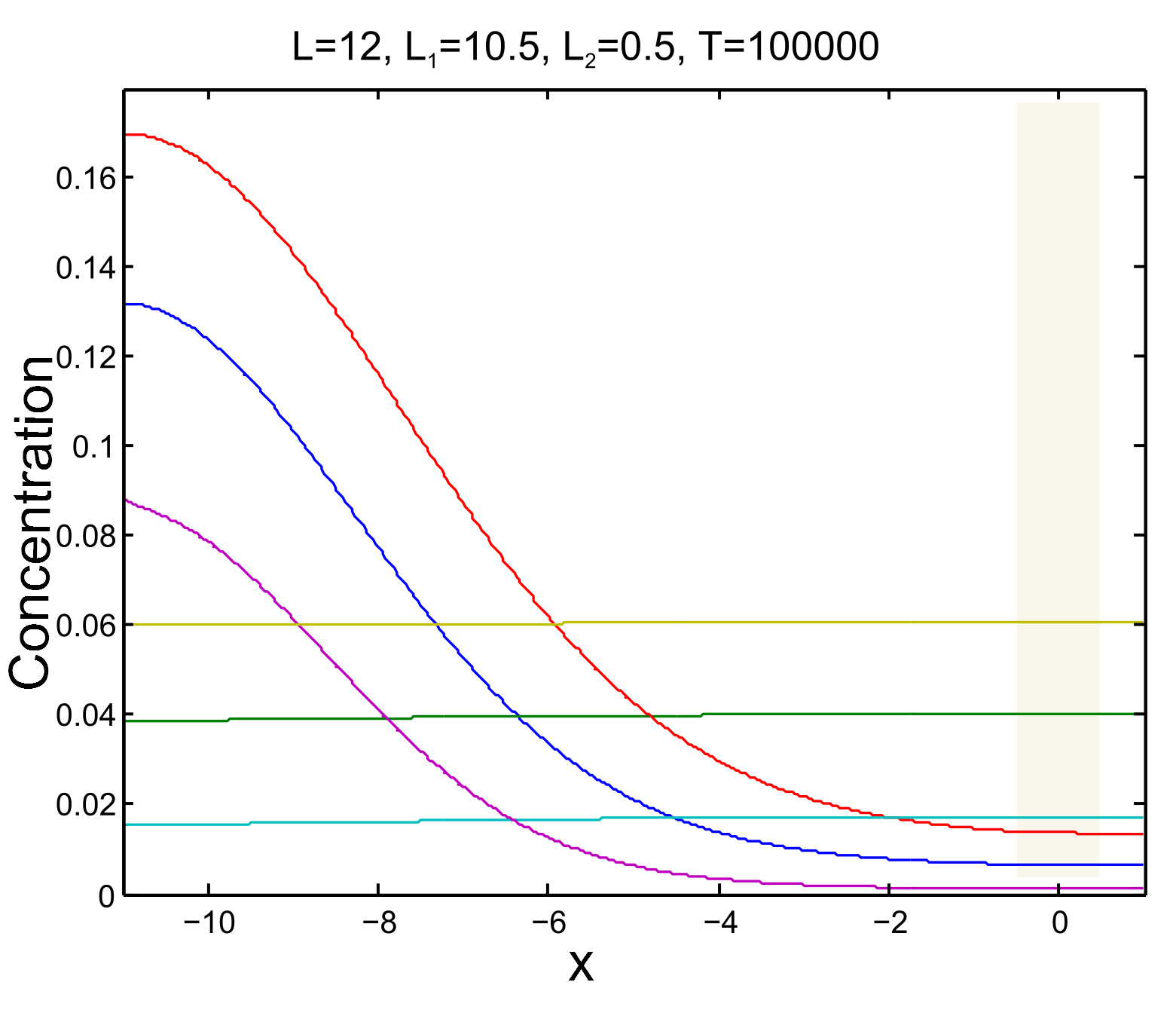}
\includegraphics[width=0.3\textwidth]{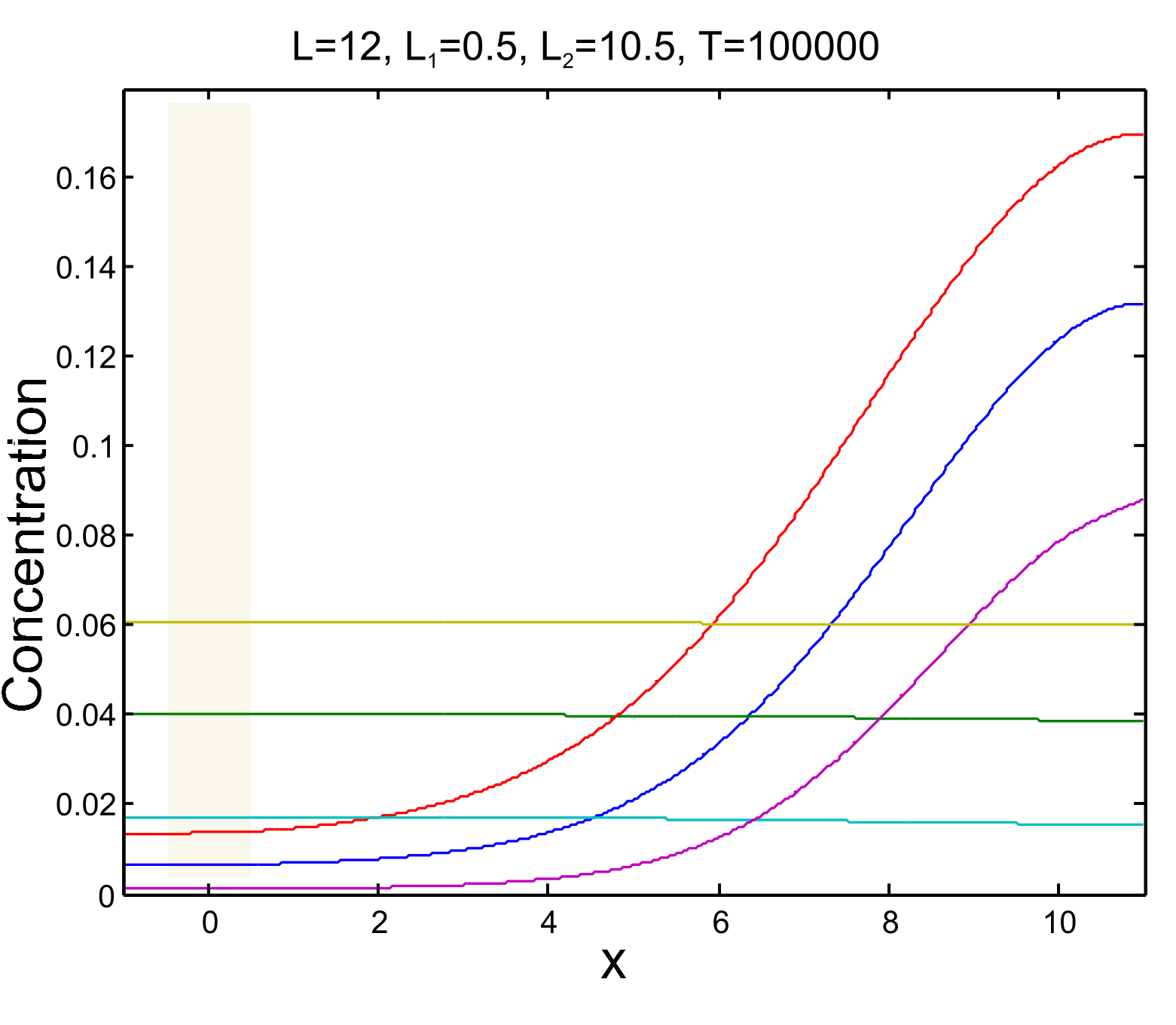} \\
\includegraphics[width=0.3\textwidth]{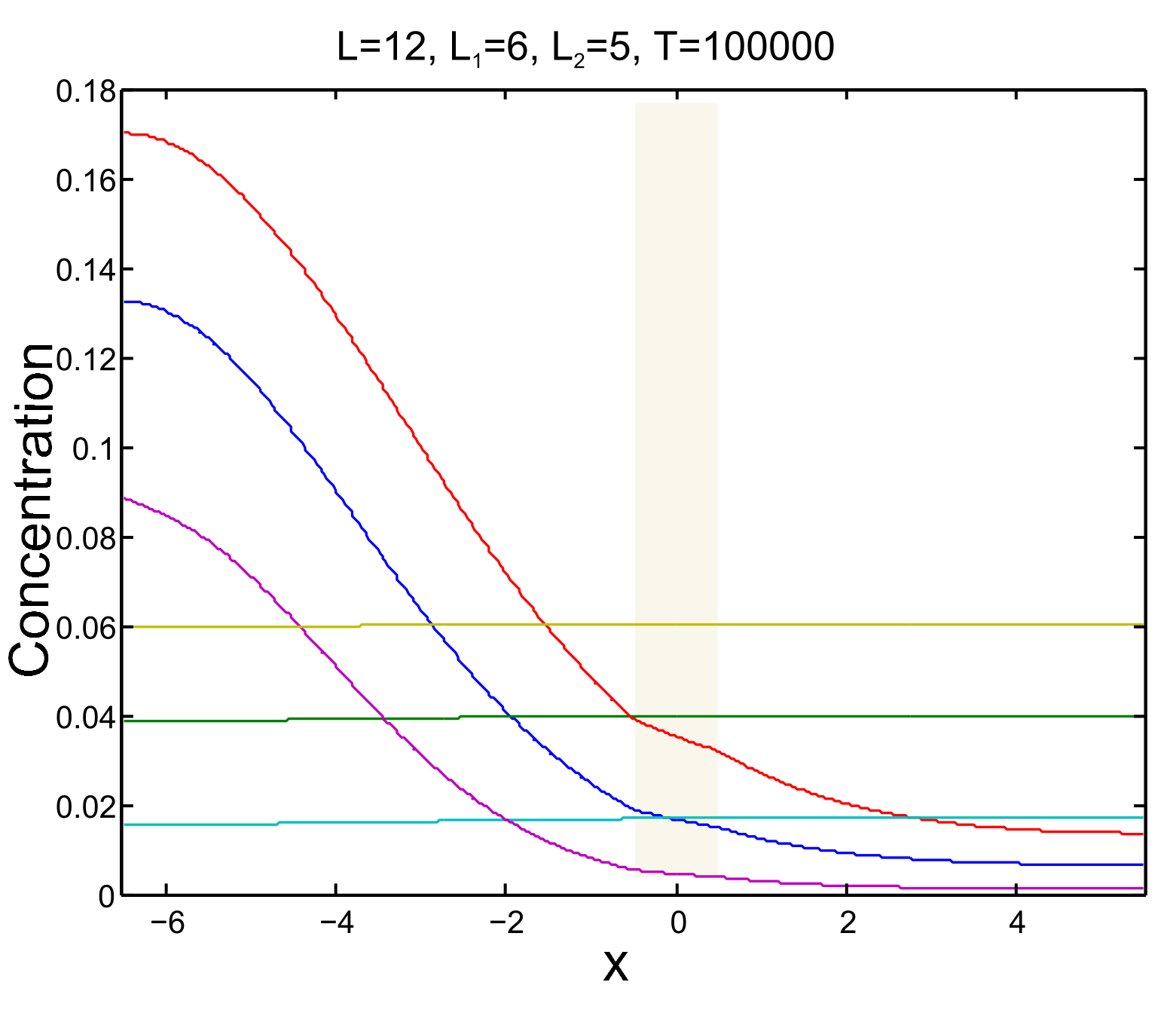}
\includegraphics[width=0.3\textwidth]{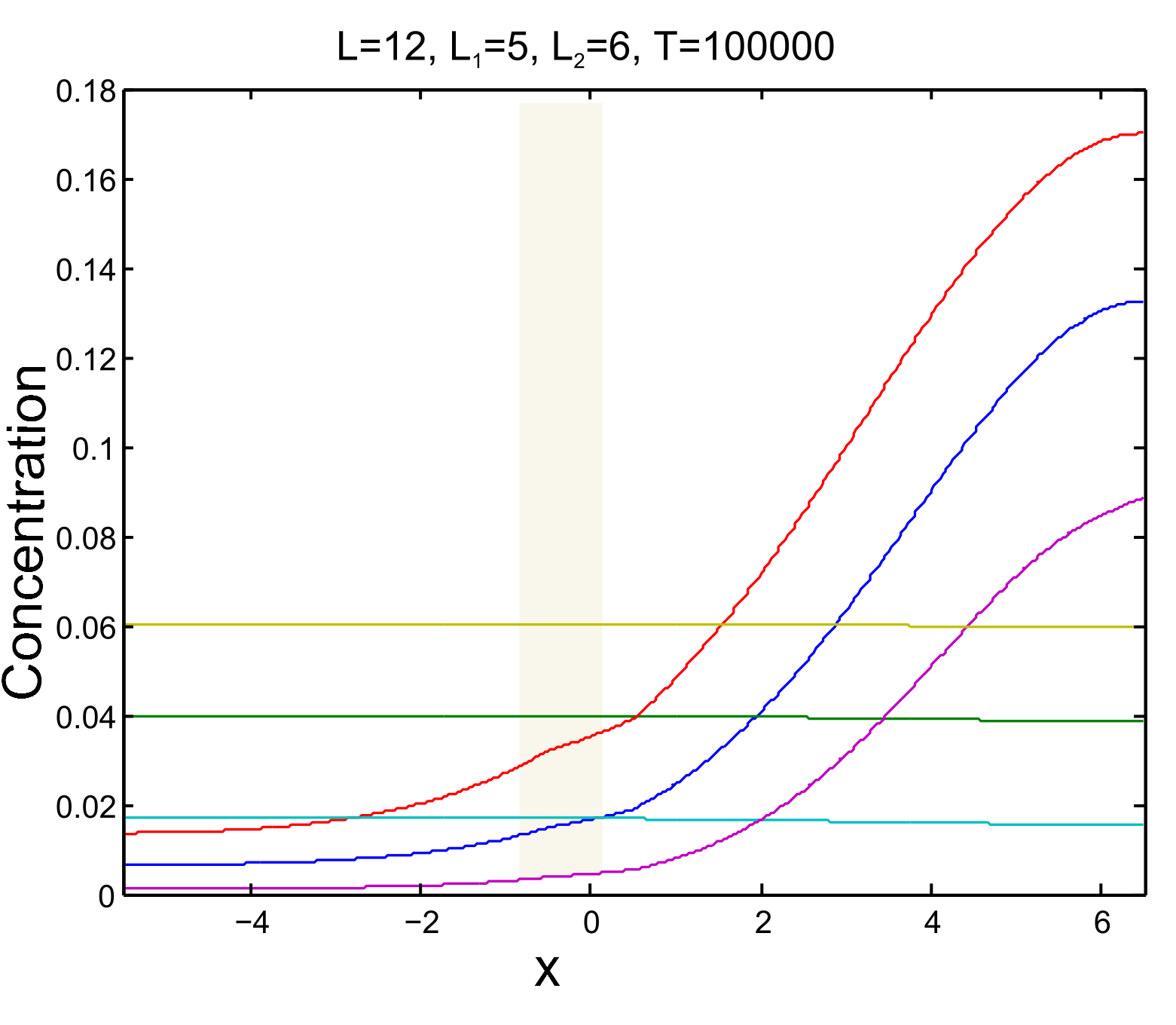}
\caption{\label{ATPolarisierung}Stationary concentrations for constant length $L$ and varying neurite lengths $L_1$ and $L_2$.}
\end{center}
\end{figure}

Some stationary solutions in the case of active transport, all obtained with the same initial value, are shown in Figure \ref{ATPolarisierung}. The polarization appears in any case in the longer neurite. 

\begin{figure}[h]
\begin{center}
\includegraphics[width=0.3\textwidth]{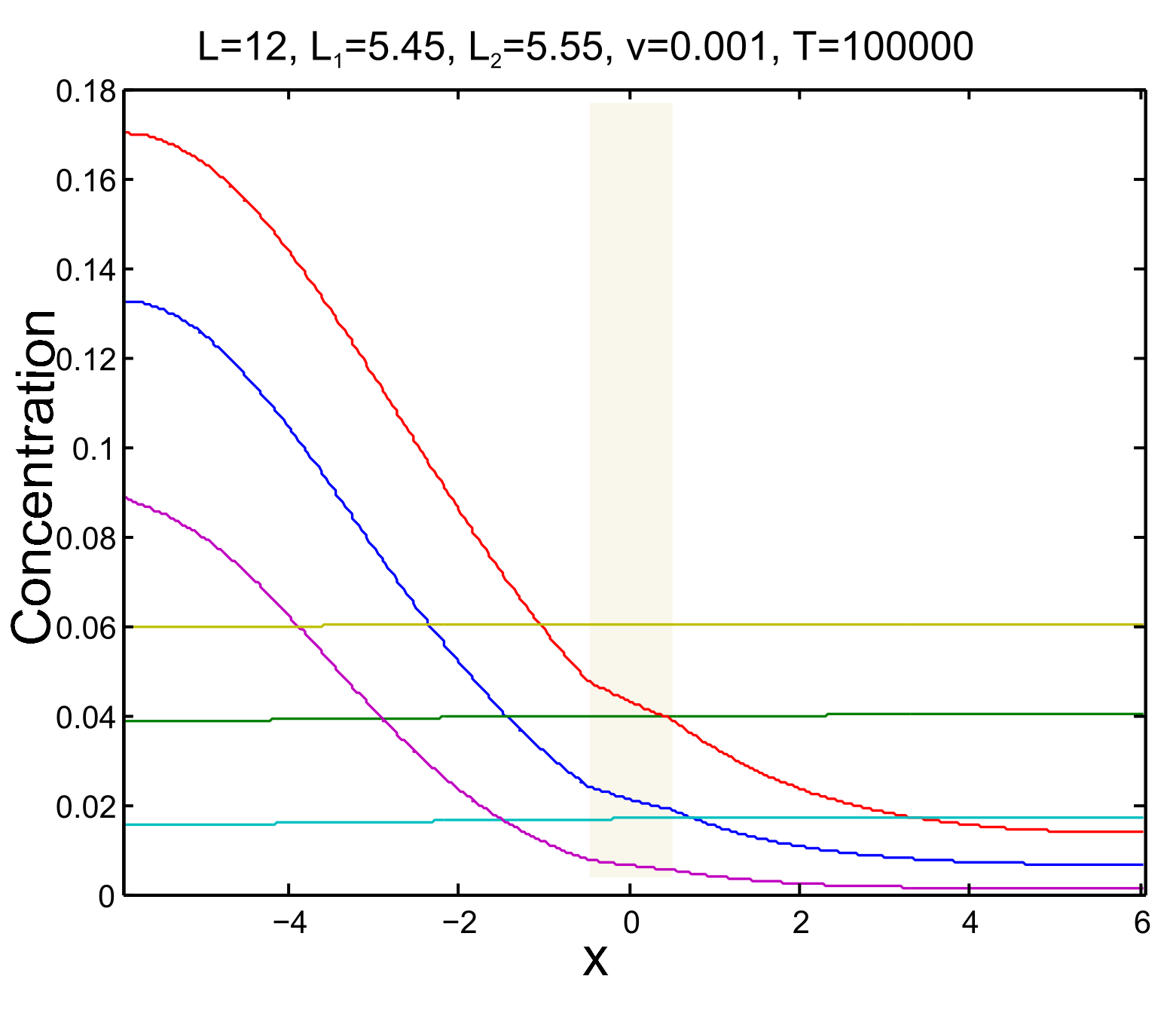}
\includegraphics[width=0.3\textwidth]{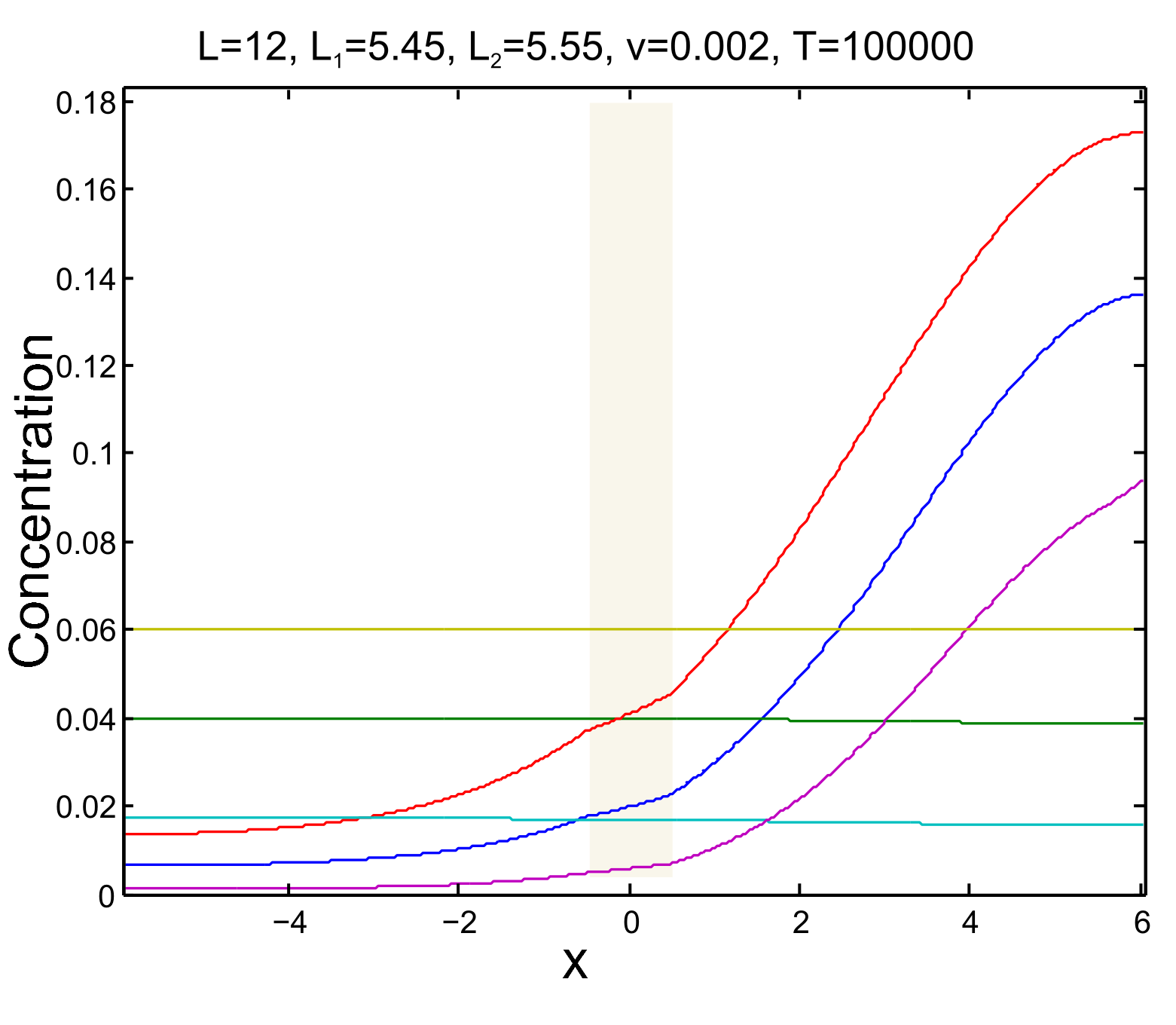}\caption{\label{ATPolarisierung2}Stationary concentrations when varying the transport speed.}
\end{center}
\end{figure}

Another interesting observation is that stronger transport always leads to polarization in the longer neurite as we illustrate by Figure \ref{ATPolarisierung2}. Here at a small transport $v=0.001$ the polarization occurs in the left neurite, which is a bit smaller. However, doubling the transport speed also yields polarization in the longer one despite the very small length difference and the positive perturbation in the shorter one.

\begin{figure}[h]
\begin{center}
\includegraphics[width=0.3\textwidth]{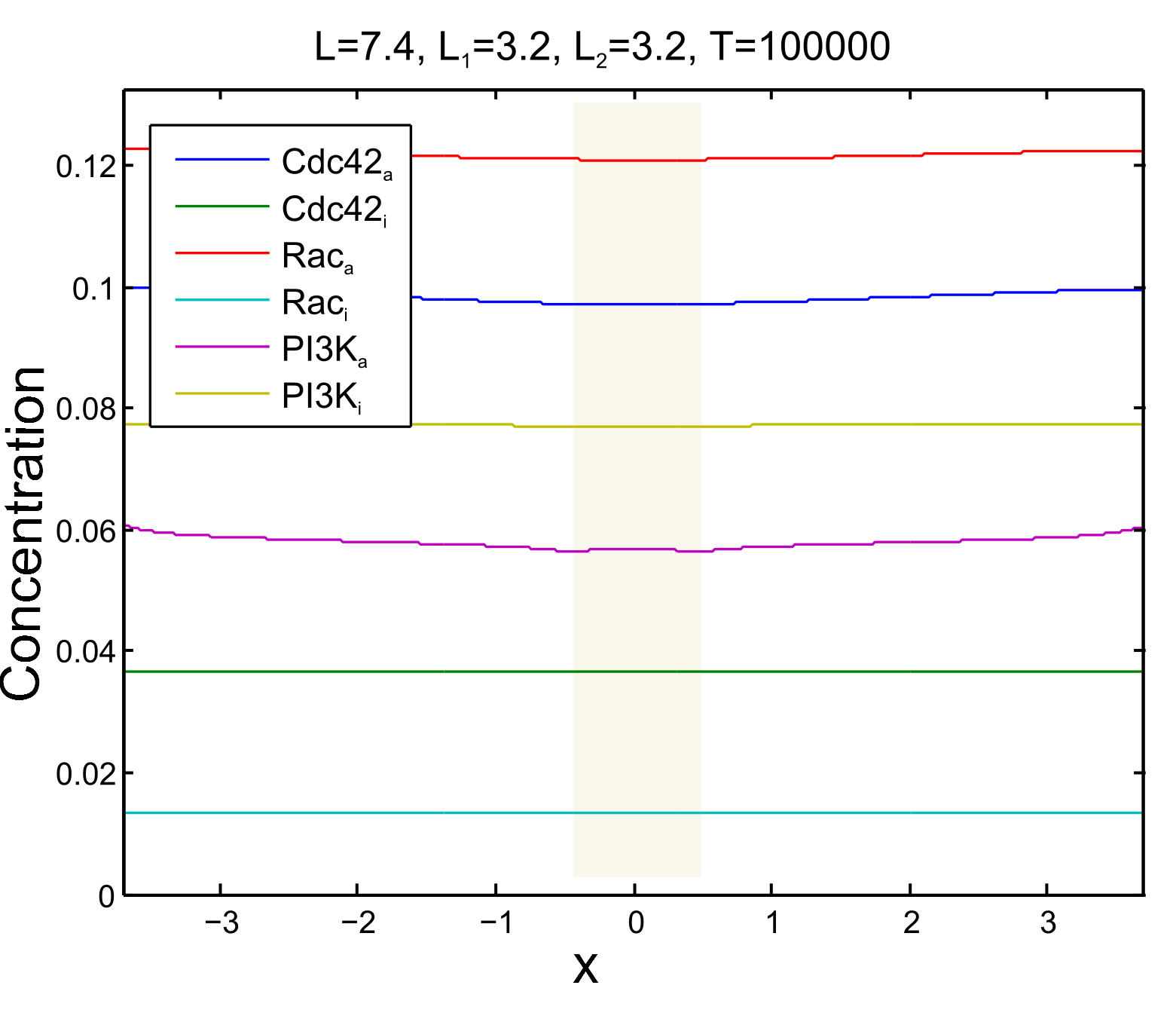}
\includegraphics[width=0.3\textwidth]{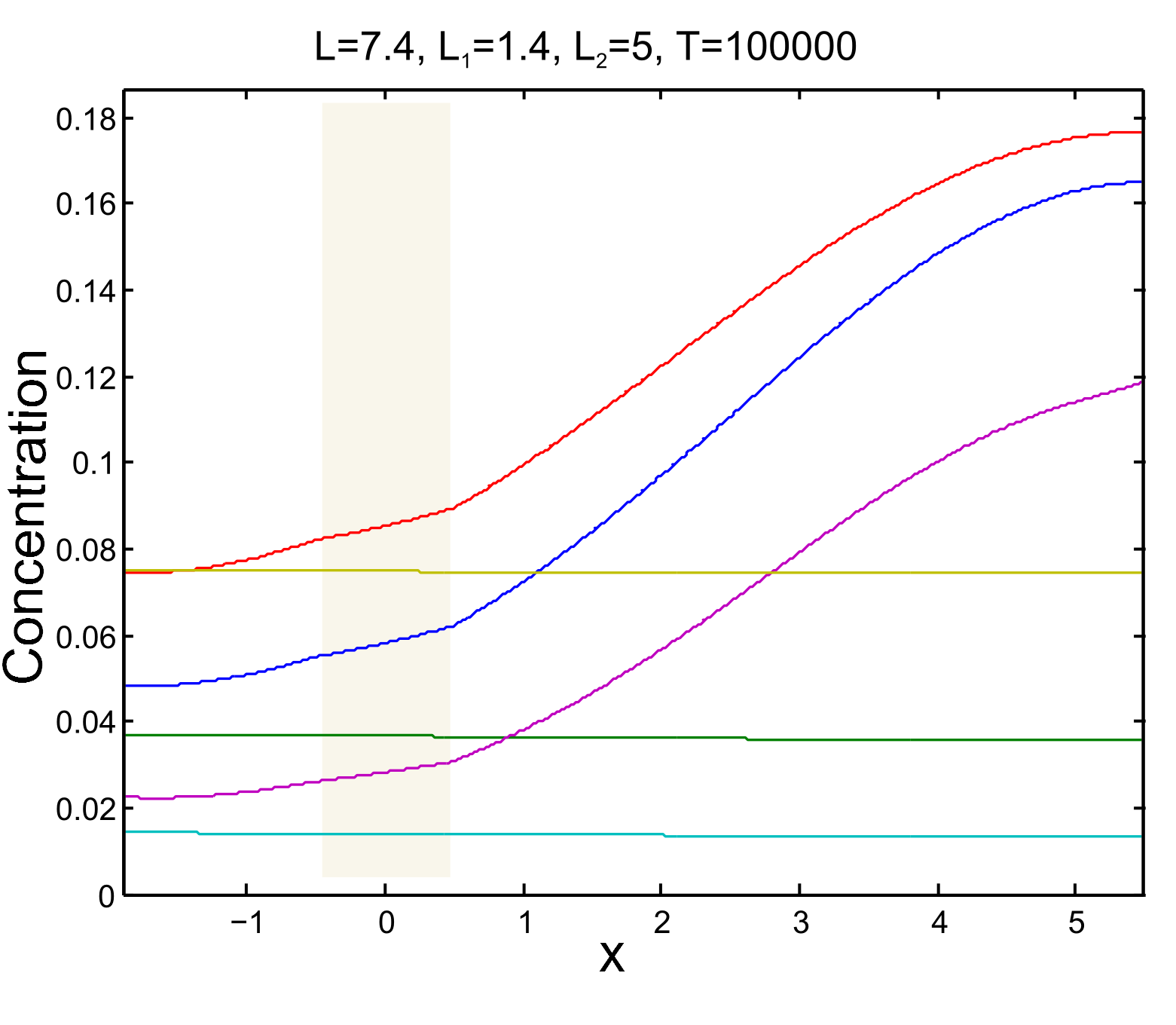}\caption{\label{ATPolarisierung3}Stationary concentrations for constant length $L$ and varying neurite lengths $L_1$ and $L_2$.}
\end{center}
\end{figure}

Finally, an experimental observation that is reproduced with active transport as well is the fact that polarization depends on the length configuration of the neurites, in particular their length difference in the case of two neurites. This is illustrated by the results in Figure \ref{ATPolarisierung3}, where the total system length $L=7.4$ is kept fixed. However, in the case of two neurites of equal length this does not suffice to produce polarization, whereas in the case of one very large and one very small neurite the polarization clearly occurs.

\section{Analysis of a Minimal Model} \label{sec:minimalmodel}

In order to gain some analytical insight into the polarization behaviour we study a minimal model in the following by reducing to the essential qualitative features, namely an active and an inactive form together with a nonlinear feedback loop in the activation.

%\begin{wrapfigure}{h}{5.5cm}
%\includegraphics[height=1.5cm]{KanonischesModell.png}
%\caption[Vereinfachtes Reaktionsschema der GTPase Rac]{\label{kanSchema}Vereinfachtes Reaktionsschema der GTPase Rac. Die Selbstaktivierung erfolgt über die Rückkopplungsschleife}
%\end{wrapfigure} \begin{align}
Denoting the active concentration by $u_1$ and the inactive by $u_2$, 
we study a system of the following form
\begin{align}
\partial_t u_1 =& \epsilon^2 \partial_{xx} u_1+\rho(\epsilon u_1)u_2-\delta u_1, \label{minimalmodel1}\\
\partial_t u_2 =& D \epsilon^2 \partial_{xx} u_2-\rho(\epsilon u_1)u_2+\delta u_1, \label{minimalmodel2}
\end{align}
for $x \in (0,1)$ and $t>0$. Clearly $\delta > 0$ and  $\rho: \mathbb{R}_+ \rightarrow \mathbb{R}_+$ is a potentially nonlinear function modelling the activation feedback loop. We normalize the mass such that
\begin{equation}
	\int_0^1 (u_1+u_2)~dx = 1. 
\end{equation}

%\subsection{The need for ultrasensitivity}

We analyze the existence and stability of stationary solutions of
\eqref{minimalmodel1}, \eqref{minimalmodel2}. Homogeneous stationary solutions satisfy 
\begin{align}
&u_2^*=1-u_1^*,\\
&\rho
(\epsilon u_1^*)(1-u_1^*)=\delta u_1^*.
\end{align}
%Assuming $\rho(0)=0$ such as in the canonical example $\rho(u_1) = u_1^2$
Stability of homogeneous solutions is given if the Jacobian of the reaction term
$$ F(u_1,u_2) = (\rho (\epsilon u_1)u_2 -\delta u_1 , 
-\rho (\epsilon u_1)u_2 +\delta u_1) $$
has eigenvalues with nonpositive real parts. The eigenvalues are given by
$\lambda_1=0$ und $\lambda_2=-\delta -\rho +\epsilon \rho^\prime u_2^*$.
Hence, we obtain the stability condition
\begin{align}
\delta +\rho (\epsilon u_1^*) \geq \epsilon \rho^\prime (\epsilon u_1^*)u_2^*.
\end{align} 
To check for Turing instability it suffices to consider the linearization of the reaction diffusion system for perturbations $\cos (k\pi)$, with a natural number $k$, since those are the eigenfunctions of the second derivative with Neumann boundary conditions. For $\mu=k^2\pi^2$ the eigenvalues of the linearizations can be computed as in the case $\mu = 0$ and we obtain the following condition for the appearance of a Turing instability:
\begin{align}
D\mu \epsilon^3 \rho^\prime u_2^* > D \mu \epsilon^2 \delta +D\mu^2 \epsilon^4 +\mu \epsilon^2 \rho .
\end{align}
We insert 
$u_2^*=\frac{\delta u_1^*}{\rho(\epsilon u_1^*)}$ and simplify for $D \gg 1$ to
\begin{align}
\frac{\rho^\prime (\epsilon u_1^*) \epsilon u_1^*}{\rho (\epsilon u_1^*)} >1+\frac{\mu}{\delta}\epsilon^2.
\end{align}
For small $L$ respectively large $\epsilon$ the $\epsilon^2$ term is dominating, thus there is no instability. For sufficiently large $L$ the 
instability condition is satisfied if 
\begin{align}
\rho^\prime(w)w>\rho(w) \quad \text{for all} \quad w>0. \label{Turingminimal}
\end{align}
Now we observe that \eqref{Turingminimal} cannot be satisfied for linear $\rho$, since this would always imply the opposite inequality for all $w$. Hence, some ultrasensitivity is needed to make the homogeneous solution unstable at some length $L$.

We also see that with a quadratic atcivation the system  can produce a length-dependent Turing instability. Consider 
\begin{align}
\rho(w)=\alpha +\beta w^2,\label{KM_Beispiel}
\end{align}
then $\rho^\prime(w)= 2\beta \epsilon^2 w$, hence instability can occur if
\begin{align}
\frac{\rho^\prime (w)w}{\rho(w)}-1=\frac{2\beta w^2}{\alpha +\beta w^2}-1 = \frac{-\alpha +\beta w^2}{\alpha+ \beta w^2}>0.
\end{align}
This is true in particular if $\alpha = 0$.

\section{Conclusions}

We have studied the mathematical modelling of neuronal polarization focusing on an experimentally observed feedback loop between PI3 kinase, Rac and Cdc42 and investigated the model behaviour by numerical solutions in a setup with two neurites separated by the soma. Our main conclusions are the following:

\begin{itemize}

\item A feedback loop in the activation of GTPases and PI3 kinase can lead to an inherent polarization mechanism at critical length.  However

\item Some kind of ultrasensitivity is needed to produce instabilities leading to polarization, the latter do not occur with simple linear reactions.

\item Symmetry breaking as observed in experiments can be obtained by active anterograde transport. This is however not specific to the transport of a particular species, and - since the mathematical form for domain growth is analogous to the active transport term - could even occur due to mechanical effects. Note the experiments of \cite{lam}, which lead to polarization caused by pulling on a neurite. 

\end{itemize}

Our results suggest a variety of future questions to be solved. Of course our model is far from being quantitative and the need for  extensions as well as calibrations of model parameters are to be clarified from experiments. The role of active transport during development is another issue that calls for further experimental investigations. 

From a mathematical point it is a challenging issue to characterize the existence of nontrivial steady states in cases of Turing instability, which so far has been successful in only few cases in literature (cf. \cite{winter1,winter2}), not for the mass-conserving model suggested here. An even more challenging and fundamental question is to understand the effect of a transport term in the equation and the detailed reasons for the symmetry breaking.

\section*{Acknowledgements}

The work of AWP has been supported by grants from the Deutsche Forschungsgemeinschaft (SFB 629 and PU102/12-1).
%%The authors thank Angela Stevens (WWU M\"unster) for links to literature.


\begin{thebibliography}{99}

\bibitem{AB00}S.S.L.Andersen, G.Bi (2000), Axon formation: a molecular model for the generation of neuronal polarity, BioEssays 22, 172-179.

\bibitem{Ar07}N.Arimura, K.Kaibuchi (2007). Neuronal Polarity: from extracellular signals to intracellular mechanisms, Nature 8, 194-205.

\bibitem{Ba09}A.P.Barnes, F.Polleux (2009), Establishment of axon-dendrite polarity in developing neurons 32, Ann. Rev. Neurosci., 347-381.

\bibitem{Ca08}F.Calderon de Anda, A.G\"artner, L.H.Tsai, C.G.Dotti (2008), Pyramidal neuron polarity axis is defined at the bipolar stage, J. Cell Sci. 121, 178-185.

\bibitem{Do88}C.G.Dotti, C.A.Sullivan, G.A.Banker GA (1988), The establishment of polarity by hippocampal neurons in culture, J. Neurosci. 8, 1454-1468.

\bibitem{EM02}S.Etienne-Manneville, A.Hall (2002), Rho-GTPases in cell biology, Nature, 420, 629-635.

\bibitem{Fi08}M.Fivaz (2008), Robust neuronal symmetry breaking by Ras-
triggered local positive feedback, Current Biol. 18, 44-50.

\bibitem{Go08}A.B.Goryachev A.V.Pokhilko (2008), Dynamics of Cdc42 network embodies  a Turing-type mechanism of yeast cell polarity. FEBS Letters 582, 1437-1443.

\bibitem{GB89}K.Goslin, G.Banker (1989), Experimental observations of the development
of polarity by hippocampal neurons in culture, J. Cell Biol. 108,  1507-1516.

\bibitem{Go05}E.E.Govek, S.E.Newey, L.E.Van Aelst (2005), The role of the Rho GTPases in neuronal development, Genes  Dev. 19, 1-49.

\bibitem{IS07}S.Ishihara, M.Otsuji, A.Mochizuki (2007), Transient and steady state of mass-conserved reaction-diffusion systems, Phys. Rev. E 75,  015203.

\bibitem{Ji07}A.Jilkine, A.F.Maree, L.Edelstein-Keshet (2007), Mathematical model for spatial segregation of
the Rho-family GTPases based on inhibitory crosstalk. Bull. Math.
Biol. 69, 1943-1978.

\bibitem{Jilkine}A.Jilkine (2009), A wave-pinning mechanism for eukaryotic cell polarization based on Rho-GTPase dynamics, PhD Thesis (UBC Vancouver).

\bibitem{JiEK11}A.Jilkine, L.Edelstein-Keshet (2011), A comparison of mathematical models for polarization of single eukaryotic cells in response to guided cues. PLOS Comp. Biol. 7, e1001121.

\bibitem{lam}P.Lamoureux, G.Ruthel, R.E.Buxbaum, S.R.Heidemann (2002), Mechanical tension can specify axonal fate in hippocampal neurons, J. Cell Biol. 159, 499-508.

\bibitem{Li09}A.Lipshtat, G.Jayaraman, J.C.He, R.Iyengar (2009), Design of versatile biochemical switches that
respond to amplitude, duration, and spatial cues. PNAS 107, 1247-
1252.

\bibitem{Me04}C.Menager, N.Arimura, Y.Fukata, K.Kaibuchi (2004), PIP3 is involved in neuronal polarization and axon formation, J. Neurochemistry 89, 109-118.

\bibitem{Mo08}Y.Mori, A.Jilkine, L.Edelstein-Keshet (2008), Wave-Pinning and cell polarity from a bistable reaction-diffusion system, Biophys. J. 94, 3684-3697.

\bibitem{Mo11}Y.Mori, A.Jilkine, L.Edelstein-Keshet (2011), Asymptotic and bifurcation analysis of a wave-based pattern formation mechanism in a model of cell polarization, SIAM J. Appl. Math. 71, 1401-1427.

\bibitem{MO10}Y.Morita, T.Ogawa (2010), Stability and bifurcation of nonconstant solutions to a reaction-diffusion system with conservation of mass,  Nonlinearity 23, 1347.

\bibitem{Ot07}M.Otsuji, S.Ishihara, C.Co, K.Kaibuchi, A.Mochizuki, S.Kuroda (2007), A mass-conserved reaction-diffusion system
captures properties of cell polarity, PLOS Comp. Biol. 3, 1040-1054.

\bibitem{Po01}M.Postma, P.J.M.Van Haastert (2001), A diffusion-translocation model for gadient sensing by chemotactic cells,  Biophys. J. 81, 1314-1323.

\bibitem{Ri01}A.J.Ridley (2001), Rho GTPases and cell migration, J. Cell Sci. 114, 2713-2722

\bibitem{Sa05}Y.Sakumura, Y.Tsukada, N.Yamamoto, S.Ishii (2005), A molecular model for axon guidance based on cross talk between Rho GTPases, Biophys. J., 89, 812-822.

\bibitem{Sa96}D.C.Samuels, (1996), The Origin of neuronal polarization: a model of axon formation. Phil. Transa. Royal Soc. Bio.
Sci. 351, 1147-1156.

\bibitem{SP04}J.C.Schwamborn, A.W.P\"uschel (2004), The sequential activity of the GTPases Rap1B and Cdc42 determines neuronal polarity. Nature Neurosci. 7, 923-929.

\bibitem{SuNa04}K.K.Subramanian, A.Narang (2004), A mechanistic model for eukaryotic gradient sensing: Spontaneous and induced phosphoinositide polarization, J. Theor. Biol. 231, 49-67.

\bibitem{Ta09}S.Tahirovic, F.Bradke (2009), Neuronal polarity, Cold Spring Harbor Perspectives in Biology 1, 1-17.

\bibitem{To06}M.Toriyama, T.Shimada, K.B.Kim, M.Mitsuba, E.Nomura, K.Katsuta, Y.Sakumura, P.Roepstorff, N.Inagaki (2006), Shootin1: a protein involved in the organization
of an asymmetric signal for neuronal polarization, J.
Cell Biol. 175, 147-157.

\bibitem{To10}M.Toriyama, Y.Sakumura, T.Shimada, S.Ishii, N.Inagaki (2010), A diffusion-based neurite length-sensing mechanism
involved in neuronal symmetry breaking. Molecular Systems Biology.
6, AN 394.


\bibitem{Wa06}M.Watabe-Uchida, E.E.Govek, L.Van Aelst, Regulators of Rho GTPases in neuronal development, J. Neurosci. 26, 10633-10635.

\bibitem{winter1}J.Wei, M.Winter (2001), Spikes for the two-dimensional Gierer-Meinhardt system: the weak coupling case, J. Nonlinear Sci. 11,  415- 458 

\bibitem{winter2}M.Winter, J.Wei (2004), Stability analysis of Turing patterns generated by the Schnakenberg model, J. Math. Biol. 49, 358- 390

\bibitem{Wi08}H.Witte H. et al. (2008), Micotubule stabilization specifies initial neuronal polarization., J. Cell Biol. 11, 619-632.

\bibitem{Yo06}T.Yoshimura, N.Arimura, K.Kaibuchi (2006), Signaling networks in neuronal polarization, J. Neurosci. 26, 10626-10630


\end{thebibliography}
\end{document}